\documentclass[12pt]{emulateapj}
      \usepackage{graphicx}
       \usepackage{natbib}
\usepackage{graphicx,rotating}
\usepackage{amssymb}
\usepackage{epstopdf}
\usepackage{amsmath}
\usepackage{mathtools}
\usepackage{natbib}
\usepackage{xcolor}
\usepackage{xspace}
\newcommand{\beq}{\begin{equation}}
\newcommand{\eeq}{\end{equation}}
\newcommand{\etal}{{\sl et~al.~}}
\newcommand{\kms}{km s$^{-1}$}

\newcommand{\FGS}{{\it FGS~}}
\newcommand{\FGSns}{{\it FGS}}
\newcommand{\hst}{{\it HST}}
\newcommand{\hsts}{{\it HST~}}

\newcommand{\HIP}{{\sl Hipparcos}}
\newcommand{\G}{{\sl Gaia}}
\newcommand{\Gs}{{\sl Gaia~}}
\newcommand{\m}{$\cal{M}$}
\newcommand{\mA}{$\cal{M}_{\rm A}$}

\newcommand{\mB}{$\cal{M}_{\rm B}$}

\newcommand{\msun}{$\cal{M}_{\odot}~$}
\newcommand{\msune}{$\cal{M}_{\odot}$}

\newcommand{\eAs}{{$\eta$\,Aql~}}
\newcommand{\eA}{{$\eta$\,Aql}}
\newcommand{\zGs}{{$\zeta$\,Gem~}}
\newcommand{\zG}{{$\zeta$\,Gem}}

\def\fdg{\hbox{$.\!\!^\circ$}}

\def\K{$Kepler$~}

\newcommand{\het}{ {\it{HET}}}
\newcommand{\hets}{ {\it{HET~}}}

\begin{document}

\received{}
\revised{}
\accepted{}

\shorttitle{The $\eta$ Aql System}
\shortauthors{Benedict \etal}

\bibliographystyle{/Active/my2}

\title{The $\eta$ Aquilae System:  Radial Velocities and Astrometry in Search of  \eAs\,B}

\author{G. Fritz Benedict} 
\affiliation{McDonald Observatory, University of Texas, Austin, TX 78712}
\author{Thomas G. Barnes III}
\affiliation{McDonald Observatory, University of Texas, Austin, TX 78712}
\author{ Nancy R. Evans}
\affiliation{Harvard-Smithsonian Center for Astrophysics, Cambridge MA 02138}
\author{William D. Cochran}
\affiliation{McDonald Observatory, University of Texas, Austin, TX 78712}
 \author{Richard I. Anderson}
 \affiliation{Institute of Physics, Laboratory of Astrophysics, \'Ecole Polytechnique F\'ed\'erale de Lausanne (EPFL), Observatoire de Sauverny, 1290 Versoix, Switzerland}
\author{Barbara E. McArthur}
\affiliation{McDonald Observatory, University of Texas, Austin, TX 78712}
\author{Thomas E. Harrison}
\affiliation{Department of Astronomy, New Mexico State University, Box 30001, MSC 4500, Las Cruces, NM 88003-8001}



\begin{abstract}
The classical Cepheid \eAs was not included in past Leavitt Law work (Benedict et al. 2007) because of a presumed complicating orbit due to a known B9.8V companion. To determine the orbit of  \eA\,B, we analyze a significant number of radial velocity measures (RV) from eight sources. With these we establish the RV variation due to Cepheid pulsation, using a twelve Fourier coefficient model, while solving for velocity offsets required to bring the RV data sets into coincidence. RV residuals provide no evidence of orbital motion, suggesting either  nearly face-on orientation or  very long period. Reanalysis of Hubble Space Telescope Fine Guidance Sensor astrometry now includes reference star parallax and proper motion priors from Gaia EDR3. As modeling confirmation, we reanalyze \zGs in parallel, deriving \zGs parallax and proper motion values consistent with Gaia EDR3, and consistent with the Benedict 2007 Leavitt Law. In an effort to further characterize \eA\,B, we hypothesize that \eAs residuals larger than those of the associated reference stars or a parallax inconsistent with EDR3 and the Benedict 2007 Leavitt Law indicate unmodeled orbital motion. Using the astrometric noise or parallax mismatch with EDR3 we estimate possible periods and mass for  \eA\,B. Ascribing photocenter motion to the photometric variation of the Cepheid, \eA\,A, yields a plausible separation, consistent with a long period, explaining the lack of RV variation. None of these approaches yields an unassailable characterization of the  \eA\,A-B system.
\end{abstract}




%

\section{Introduction}

Cepheids are 
prime objects for distance determination using a Period-Luminosity Relation, now known as the
Leavitt Law (hereafter, LL). To confirm our understanding of 
pulsation, identifying
those in binary or multiple systems from which
masses can be measured also has high value.

$\eta$ Aql was in fact the first Cepheid in which 
light variation was discovered, by Pigott in 1784, 
having  preceded the discovery of variation in 
$\delta$ Cep itself by only weeks.  It has 
comparatively well-behaved variation with a period
of 7.18$^d$. 

An early observation with the 
International Ultraviolet Explorer (IUE) satellite 
found not the low flux of a cool Cepheid, but a 
strong ultraviolet flux from a hot star \citep{Mar80}, surprising since a series of 
radial velocity studies in the 20th century showed no clear indication of orbital motion.  
UV spectrophotometry of \eAs\,B \citep[table 8,][]{Eva91} further confirmed the connection  of \eA\,B with \eA\,A. 
That result, if components A and B have the same distance from us, indicated an absolute magnitude range for \eA\,A, $-3.39<M_V<-3.74$ (depending on  IUE spectrum wavelength and/or which model atmosphere used), which with $V-K=1.89$, yields $-5.28<M_K<-5.63$, agreeing with the  \cite{Ben07} LL within the \cite{Eva91} range. This further supports the assertion that components A and B are NOT a chance alignment.

$\eta$ Aql was observed by Hubble Space Telescope 
(\hst) Fine Guidance Sensor (\FGSns) in a program
to determine Cepheid parallaxes \citep{Ben07}. Unfortunately, it was the one star 
in that program for 
which a parallax could not be determined because 
of perturbations thought to be from binary motion. 
But, with no information about an orbit with which to model
them, \eAs remained without an \FGS parallax.

Using the Wide Field Camera 3 (WFC3) on \hst,
\cite{Eva13} found a companion 0.66$\arcsec$ 
from the Cepheid.  Subsequently, \cite{Gal14} 
found the same companion with the ESO VLT NACO instrument.
They estimated the unreddened H magnitude to be 
9.34 $\pm$ 0.04, corresponding to the spectral type 
range F1 V to F6 V.  

 These results combine to present the following picture 
of the $\eta$ Aql system.  In addition to the Cepheid, 
it contains a resolved companion with an F spectral type (component C) and a hot third star (component B) closer to the 
Cepheid with a spectral type of B9.8V \citep{Eva91}.   


We note that there has recently been an interest in the 
occurrence of triple systems as well as binaries.  The 
distribution of parameters of binary and higher multiples 
(mass ratio, separation, eccentricity) are important constraints
in modeling star formation.  In addition, the interaction between
components of a triple system via the Kozai-Lidov mechanism (e.g. \nocite{Nao16} Naoz 2016)
plays an important role in post-formation evolution of the system.  
 
 We revisit \eA, now with  new high-precision radial velocities (RV) 
from the Hobby-Eberly Telescope (HET), the Harlan J Smith telescope of McDonald Observatory, the Hermes Spectrograph on the Mercator Telescope, and the Coralie spectrograph on the Euler telescope. Our goals included measuring a parallax for \eA, establishing a perturbation orbit for  \eAs A (assisted by orbit information derived from RV),  and obtaining a dynamical mass for the hot companion, \eAs B. We first present how we obtained (Section~\ref{RVd}) and reduced (Section~\ref{RVr}) our new, high-precision RV data. We next derive  a Fourier description of the RV pulsation signal, and fail to identify the signature of an \eAs B perturbation orbit in the residuals to that pulsation model (Section~\ref{RVf}). Section~\ref{Ap1} presents our analysis of the \hsts astrometry yielding a parallax for \eAs and weak evidence for an astrometric perturbation. We  place \eAs on the \cite{Ben07} LL, estimate an orbital period range for the B9.8V companion, discuss remaining issues in Section~\ref{Disc} and summarize in Section~\ref{Summ}. 

For the astrometry we abbreviate millisecond of arc, mas, and transform Julian Day to a truncated Julian Date, TJD=JD-2400000.

\section{Radial Velocities} 
\subsection{Pulsational Phase}
\eAs has an evolving pulsational period \nocite{Ber00,Eng15} (Bernikov, 2000; updated based on Engle, 2015) which complicates computation of the pulsation phase.
Phase can be calculated from
\beq
\rm{HJD}_{\rm pred} = 2411999.693 + 7.17654682 E + (2.90x10^{-8}) E^2 \label{QUADph}
\eeq
where HJD$_{\rm pred}$ is the  Heliocentric Julian Day  time of photometric maximum, 
and
E  is the cycle count from 2411999.693. We obtain phase by solving quadratic Equation~\ref{QUADph} for each observed HJD. The fractional part of the solution is the phase, $\Phi$. 

\subsection{The RV Data}\label{RVd}
We interpret the failure of the previous parallax determination \citep{Ben07} as due to a nearly face-on orbit for the $\eta$ Aql AB system. This created degeneracy in the parallax solution between parallactic motion and orbital motion. Without knowledge of the orbit, and only six distinct epochs of astrometry, \cite{Ben07} could not determine a perturbation orbit and thus a parallax.
 
$\eta$ Aql, known to be a triple system, has a wide companion much too distant from the Cepheid to cause angular motion on a few years timescale, the duration of the astrometric observations.  High--precision radial velocities of the primary are vital to disentangle the characteristics of the inner system.  Cepheid velocity curves repeat very precisely, and techniques have been developed to remove the pulsation velocity from orbital motion with high accuracy \citep{Eva00}. These techniques have success because the orbital period  is typically much longer than the typical Cepheid pulsation period.
 
Given estimates for the masses of \eAs A and B (5.7, 2.3\msune, Evans et al. 2015) and the absence of any detectable periodic RV variation with amplitude larger than 1 \kms, to discover and measure the radial velocity variation of the inner binary $\eta$ Aql AB requires that we determine a very high quality pulsation curve for $\eta$ Aql, and measure a time drift in the center of mass velocity relative to that curve caused by the orbital motion. With the Hobby-Eberly Telescope High Resolution Spectrograph \citep{Tul98} and the 2.7m Harlan J Smith telescope coude spectrograph (hereafter \hets and HJS respectively), both using an iodine cell \citep{Coc04}, we should do better than previous ground-based efforts. (We used the 9.2 m \hets for its queue--scheduled capability, not its aperture.) The uncertainties in the iodine cell radial velocities are sufficiently small to detect orbital motion on the scale of a few tens of m/s. In a very short integration time of 35 sec we typically obtained S/N $\sim 600$/pixel for the HET. 
 
We received \het-HRS time in four scheduled trimesters, totaling 111 observations over 107 nights in the interval 10 August 2010 -- 16 November 2011.  We used the 316g cross disperser, centered at 5936{\AA}, with the echelle grating in the ``central'' position (i.e. on the blaze).  The spectrograph was fed with a fiber that subtended 2 arcsec on the sky. We used a spectral resolving power of 60,000 for all of the I$_2$ cell observations.  The observing sequence included the standard calibration frames shared by all programs using a given \het-HRS configuration:  5 bias frames, 11 flats without the I$_2$ cell, 3 flats with the I$_2$ cell and 1-2 Th-Ar frames for wavelength calibration, and then one frame for $\eta$~Aql with the I$_2$ cell at some time during the night.  On three different nights we observed $\eta$~Aql at high signal/noise at resolving power of 120,000 with and without the I$_2$ cell to obtain a stellar ``template'' spectrum for the high-precision radial velocity computation.  These template spectra were obtained at pulsation phases of 0.39265, 0.65822 and 0.21875.   Thus, they sample very different phases of the pulsation cycle. 

Observations with the McDonald Observatory 2.7m Harlan J Smith Telescope were obtained using the Tull Coude Spectrograph  \citep{Tul95}.  This instrument is a cross-dispersed white-pupil echelle spectrograph covering 3750 to 10200 \AA.   A 1.2 arcsec entrance slit gave spectral resolving power of 60,000.   We used a temperature-controlled $I_2$ vapor absorption cell in front of the spectrograph slit to provide extremely stable wavelength calibration and to enable excellent reconstruction of the instrumental profile.

We  obtained a total of 50 spectra of \eAs with the HJS Tull Spectrograh  and I2 cell between 2018 September 15 and 2019 September 30.  We also obtained spectra of \eAs without the I2 cell on two separate nights for template spectra for the precise radial velocity calculations.  These were obtained at eta Aql pulsation phases of 0.07198 and 0.39497.  A 50 sec integration with the HJS Tull coude spectrograph gives S/N of $\sim$300/pixel.

\subsection{\hets and HJS Data} \label{RVr}
The data from each night of  \hets and HJS observations were reduced separately.  We used  automated scripts of IRAF tasks to perform standard bias removal, scattered light removal and flat-fielding of the data frames.  The apertures for each echelle spectral spectral order were traced, and the spectra were extracted.  Wavelength calibration was obtained from the Th-Ar hollow-cathode lamp spectra.  Since the \het-HRS did not contain an exposure meter, the mid-exposure time was estimated to be half way between the exposure start and stop time. The HJS Tull spectrograph does contains an exposure meter which records the relative flux passing through the entrance slit in 1-second intervals as a function of time.  We used this time series to compute the barycentric correction for each exposure meter time interval according to the procedure of \cite{Wri14}.  We then weighted each of these by the exposure meter flux to compute the flux-weighted barycentric correction for the exposure.

Our  primary goal is to establish an RV orbit for the perturbation caused by the B9.8V companion, and incorporate that orbit into the astrometric analysis, permitting a determination of both a parallax and companion mass. We first model the RV variation due to Cepheid pulsation, then search for RV variation (an orbit) in the residuals to that determination.   We computed radial velocities from the observed \hets and HJS spectra using a method very similar to that described by 
\cite{Mar92}. 
The radial velocity code computes the shift of the template spectrum relative to the I2 spectrum that is needed  to match the observed program spectrum, after convolution with the model instrumental profile, as defined primarily by the observed shape of the superimposed I$_2$ absorption lines.  The ``radial velocity'' computed is just $\mathrm{c}  \times (\delta\lambda/\lambda)$ where $\delta\lambda$ is the observed spectral shift. The apparent radial velocity variations due to the Cepheid pulsation result from the time variations in the overall photospheric velocity field due to the star's periodic expansion and contraction, rather from true center-of-mass motions of the star.  

\begin{center}
\begin{figure}
\includegraphics[width=6in]{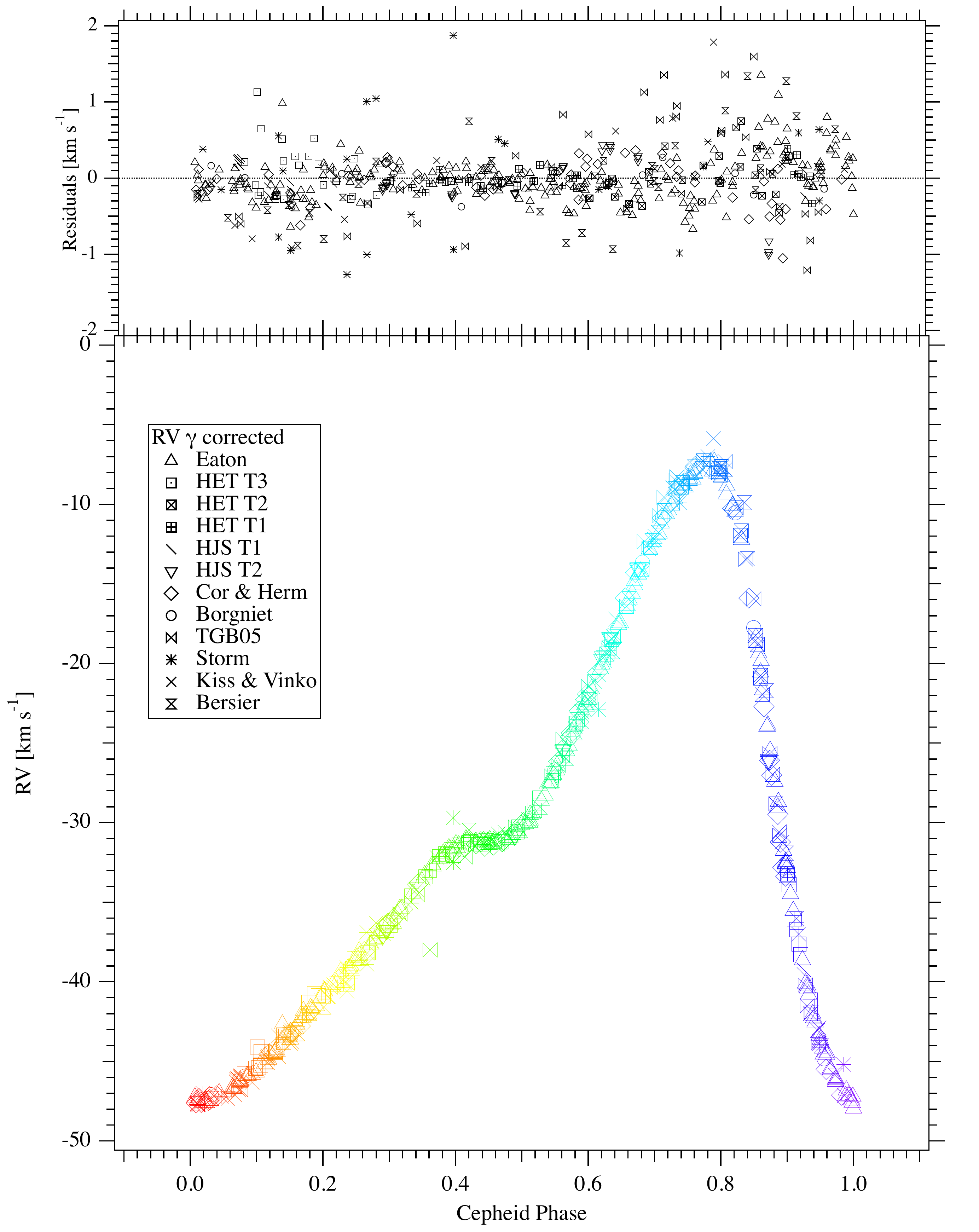}
\caption{The $\gamma$ corrected RV ($\gamma$ corrections produced by the Fourier modeling of pulsational RV, coefficients listed in Table~\ref{tbl-Fou}), plotted as a function of quadratic phase (Equation ~\ref{QUADph}). We list the RV sources and associated $\gamma$ values  in Table~\ref{tbl-RVs}. We color coded the RV values by Cepheid phase (red=0 to blue=1). Residuals are in the upper panel. 
}
\label{fig-RVph}
\end{figure}
\end{center}

The stellar line profiles undergo significant shape changes through the Cepheid pulsation cycle.  Indeed, lines formed at different depths in the atmosphere will exhibit different profile variation patterns.  The "radial velocity" measurement process basically computes the first moment of the mean stellar absorption line profile.   For a non-pulsating and inactive star, this would be an excellent approximation to the center-of-mass motion of the star. Since the  different template spectra were obtained at different Cepheid pulsation phases, they have different stellar line profile shapes.  Thus, one might expect subtle variations in the RV curves computed with the different templates.   The \hets template with quadratic phase, $\Phi=0.39$, yielded velocities with smaller internal errors, possibly because of proximity in phase to 
structure in the pulsational RV (e.g., Figure~\ref{fig-RVph}). Superimposed on these line profile variations is the true RV orbital motion of stellar components A and B. Our recent \hets velocities are given in Table~\ref{tbl-HET}, along with phases calculated from Equation 1, and template identification.
Our recent HJS velocities are given in Table~\ref{tbl-HJS}, along with phases calculated from Equation 1, and template identification.

\subsection{Other RV Data}
 We also had access to 217 radial velocities of $\eta$ Aql from the 2m Tennessee State University Automatic Spectroscopic Telescope \citep{Eat07, Eat20} 
obtained over a period of two years: late 2007 to late 2009. 
Table~\ref{tbl-EAT} contains the \cite{Eat20} velocities kindly sent in advance of publication. Table~\ref{tbl-RIA} contains an additional unpublished 55 high-precision RV from the Hermes spectrograph on the Mercator Telescope (La Palma) and from the 'Coralie' spectrograph on the Euler Telescope at La Silla.. See \cite{And16b} for an example of reduction details.

\subsection{Fourier Decomposition of Pulsation-induced RV}\label{RVf}

We seek to minimize residuals to a Fourier coefficient derived RV=f(phase) description of the observed RV variation \citep{Eva15}. We use GaussFit \citep{Jef88} with these equations of condition;
\beq
{\rm RVC} = V(Phase) - V(obs) - \gamma(S)
\eeq
and
\beq
         V(Phase) = \gamma(S) + \sum_{\rm{j=1}}^{N}(a[j]*cos(j*z) + b[j]*sin(j*z))
\eeq
where RVC is the Cepheid pulsation signature, V(Phase) is calculated from the Fourier coefficients, a and b;  $z=2\pi*$Phase;  V(obs) are the measured velocities from each source; and $\gamma(S)$ is a velocity offset which depends on
the RV source (listed in Table~\ref{tbl-RVs}). We modeled the velocities (sources in Table~\ref{tbl-RVs}) with N = 6, 10, 12, and 14 coefficients and found a $\chi^2$ minimum at 12 coefficients. We list the coefficients with error estimates in Table~\ref{tbl-Fou}. Figure~\ref{fig-RVph} contains (top) the residuals from a 12 coefficient modeling of  pulsational RV for  all the RV sources listed in Table~\ref{tbl-RVs}, and (bottom) all velocities corrected for the $\gamma$ offsets listed in Table~\ref{tbl-RVs}. 

\subsection{No RV Perturbation Orbit}
\begin{center}
\begin{figure}
\includegraphics[width=6in]{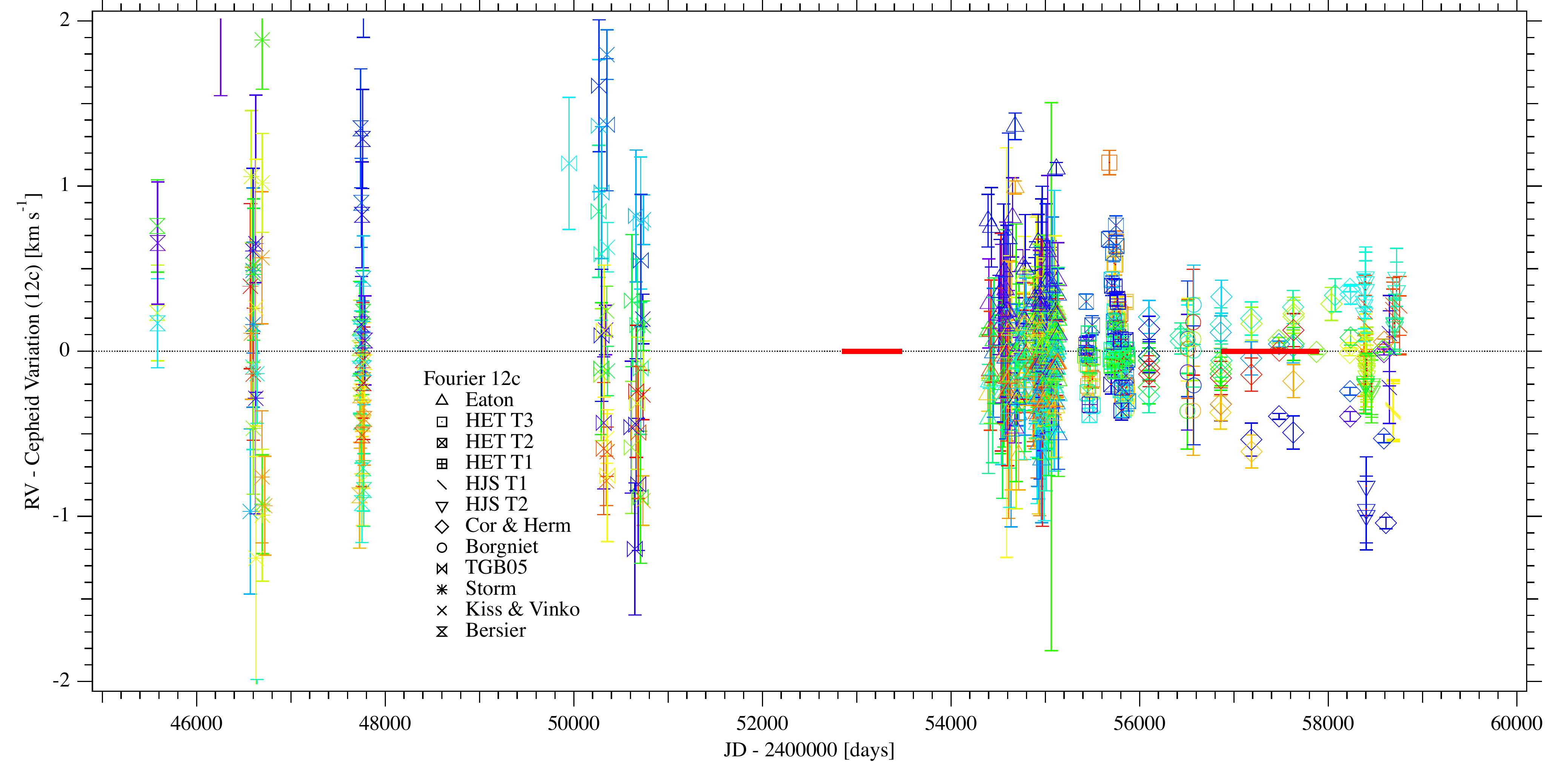}
\caption{Residuals (Figure~\ref{fig-RVph}, top) to the pulsation-induced RV, modeled by the Fourier coefficients listed in Table~\ref{tbl-Fou} plotted as a function of time for the sources listed in Table~\ref{tbl-RVs}. Errors come from Tables~\ref{tbl-HET},~\ref{tbl-HJS},~\ref{tbl-EAT},~\ref{tbl-RIA}, and the various published sources in Table~\ref{tbl-RVs}.  RV values are color coded by Cepheid phase as for Figure~\ref{fig-RVph} (red=0 to blue=1). The solid red bars 
on the zero RV axis indicate astrometric  coverage from {\it HST}/\FGS (left) and \Gs EDR3 (right).}
\label{fig-RVt}
\end{figure}
\end{center}

\begin{center}
\begin{figure}
\includegraphics[width=6in]{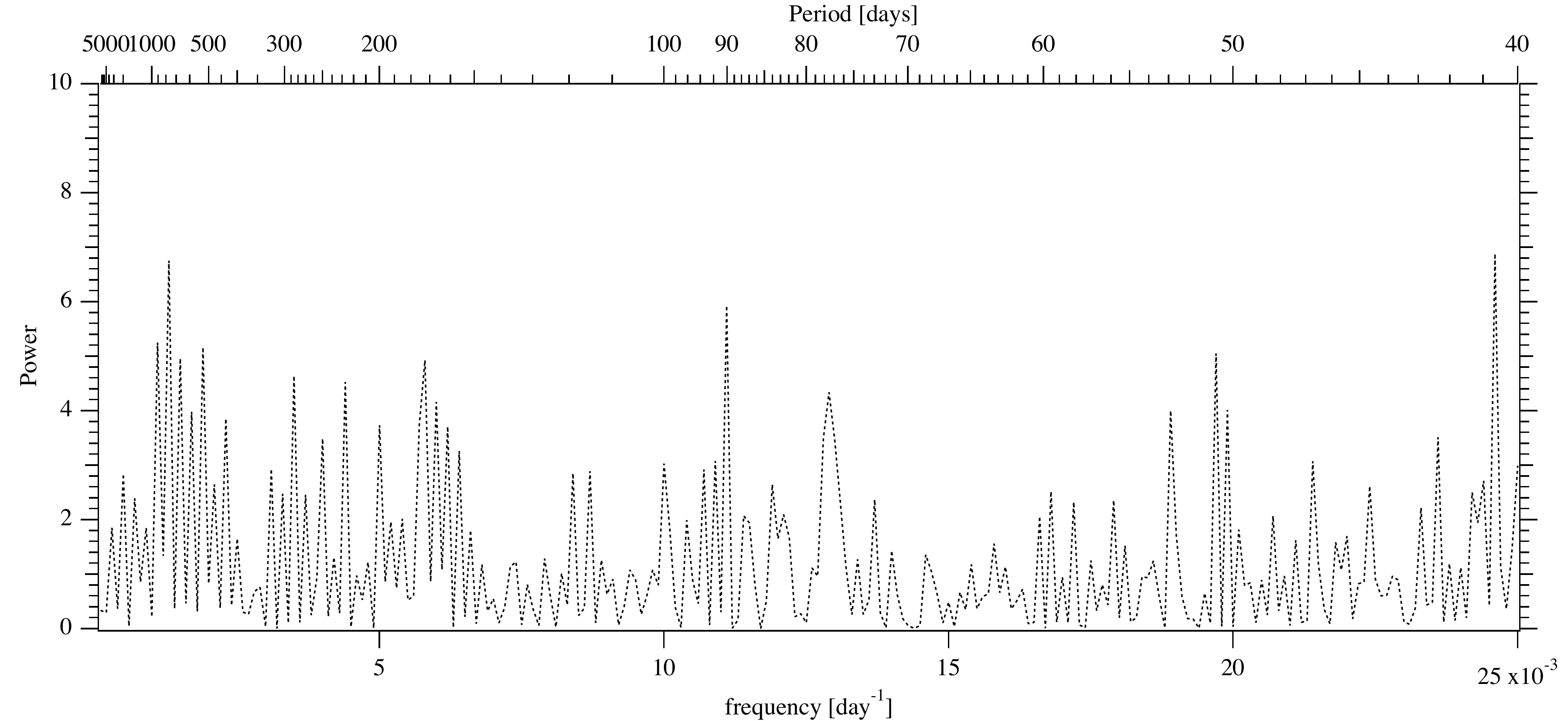}
\caption{Lomb-Scargle periodogram of Figure~\ref{fig-RVt} RV. No peak indicates a false alarm probability less than 61\%. The minimum frequency range (maximum period) explored corresponds to roughly 2/3 the total time span of the RV data.}
\label{fig-RVLS}
\end{figure}
\end{center}

Once the RV signature due to pulsation has been removed, we are left with the residuals to the Fourier fits to the various
Table~\ref{tbl-RVs} velocity sources. We search these for a perturbation caused by the B9.8V companion. Figure~\ref{fig-RVt} plots those residuals against TJD. Even though the average internal errors for many of the Table~\ref{tbl-RVs} investigations are small, we note quite large scatter in the residuals for all  data sets. We ascribe this scatter to the assumption in the RV computation processes that the stellar line profiles do not change with time. Visual inspection fails to provide any obvious RV variation due to orbital motion. Figure~\ref{fig-RVLS} shows a Lomb-Scargle periodogram of the RV residuals from the top of Figure~\ref{fig-RVph}. The lack of any significant peak at any of the probed periods implies either a nearly face-on orbit with an as yet unknown period or a a period significantly larger than those probed in Figure~\ref{fig-RVLS}.

\section{\hst / \FGS Astrometry}  \label{Ap1}

An absolute $K-$band magnitude for \eAs is  a major goal of the present investigation.  To place \eAs (with log$P$=0.85592) exactly on the \cite{Ben07} LL, requires a $K-$band absolute magnitude (with  absorption, $A_K = 0.05$ mag), $M_K=-5.23$. This in turn requires a parallax, $\varpi_{\rm abs} = 3.77$ mas with no Lutz-Kelker-Hanson (LKH) bias correction\footnote{Once a measured parallax exists we apply the LKH correction, explained in section 5 of \cite{Ben07}.}.
 \cite{Ben07} were unable to derive an \eAs parallax consistent with that  LL. At that time we blamed this failure on an unmodeled orbital perturbation due to the B9.8V companion, \eA\,B. We now revisit the \eAs astrometry, and to validate our newer astrometric modeling and explore the odd behavior of \eA, we use another Cepheid, \zGs with log$P$=1.00649, as a control sample. The \zGs EDR3 parallax (Table~\ref{tbl-1}) yields (assuming the Benedict et al. 2007 $A_K=0.02$) $M_K=-5.60\pm0.15$ mag, in agreement with the \cite{Ben07} prediction for that log$P$, $M_K=-5.73$.

\subsection{The Astrometric Data}
\begin{center}
\begin{figure}
\includegraphics[width=4in]{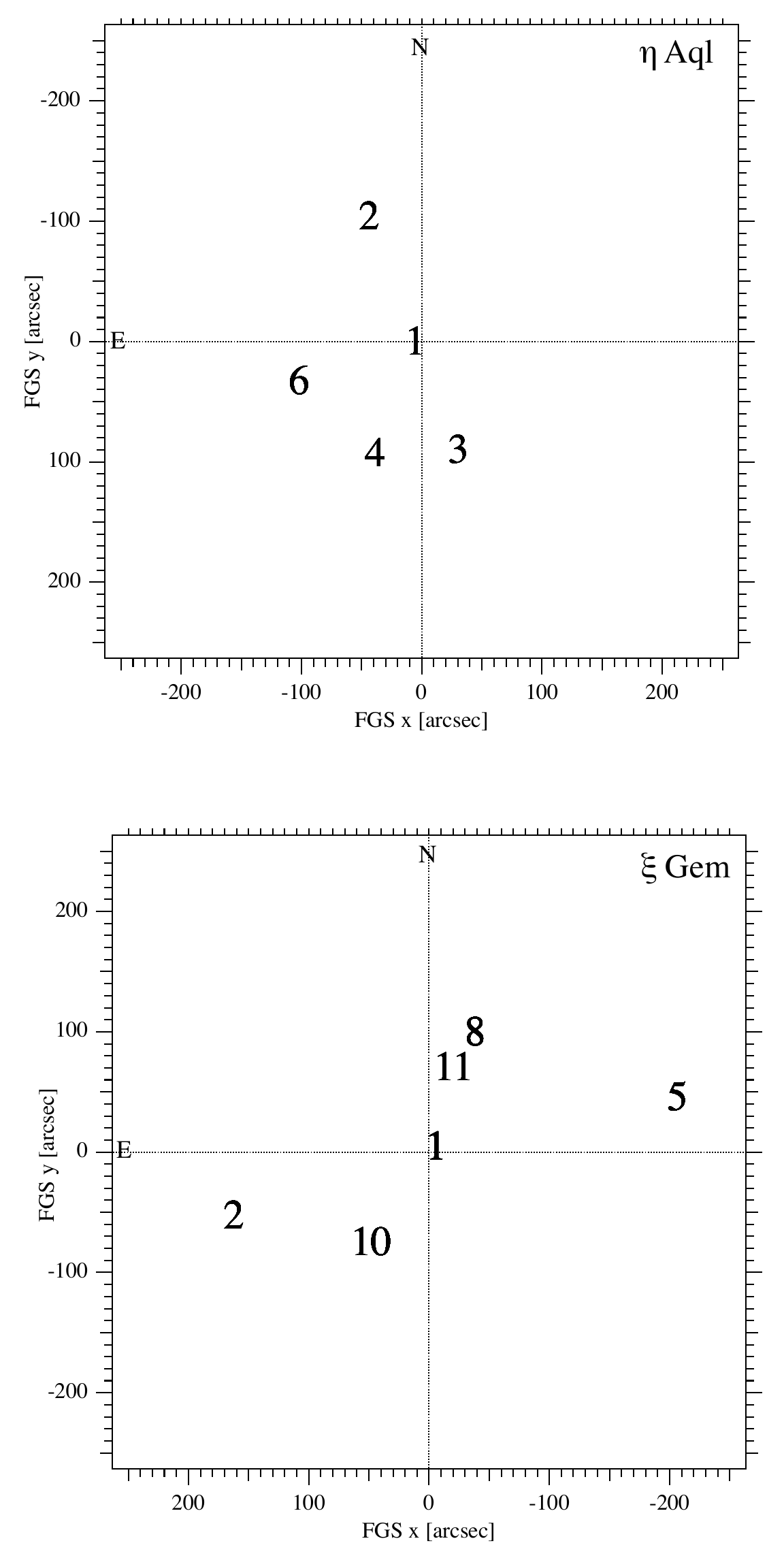}
\caption{Top: relative positions of \eAs (1, center) and the astrometric reference stars (2,3,4,6). Bottom: relative positions of \zGs (1, center) and the astrometric reference stars (2,5,8,10,11).  Table~\ref{tbl-1} identifies each reference star.}
\label{fig-Find}
\end{figure}
\end{center}

For \eAs the astrometric data consist of ten discrete sets containing a total of 111 positions,  37 of \eA, and 74 for four reference stars, all secured  with \FGSns\,1. One entire set of \eAs observations was discarded because of anomalously large residuals. Inspection of the processed data indicated poor drift correction\footnote{See \cite{Ben98}, section 3.3.2} due to excessive spacecraft motion. The reference star average positional error in $y$ has a typical value of 2.6 mas. For this field the reference star average error in $x$ has an atypical value, 6.2 mas. The \zGs data  consist of eleven discrete sets containing 201 positions, 51 of \zG, and 150 for five reference stars. In contrast to \eA, the $\zeta$ Gem field \citep{Ben07} has reference star average $x,y$ errors of 2.4 and 3.7 mas, respectively.  Relative positions of the astrometric reference stars are plotted in Figure~\ref{fig-Find}, with \Gs EDR3 identifications; parallaxes, $\varpi$; proper motions, $\mu_{\alpha}$, $\mu_{\delta}$; and $G$ magnitudes listed in Table~\ref{tbl-1}.  At each epoch we measured each reference star 1 -- 3 times, and the Cepheid 3--4 times. 


\subsubsection{Modeling Priors}\label{CORR}
\begin{figure}
\includegraphics[width=7in]{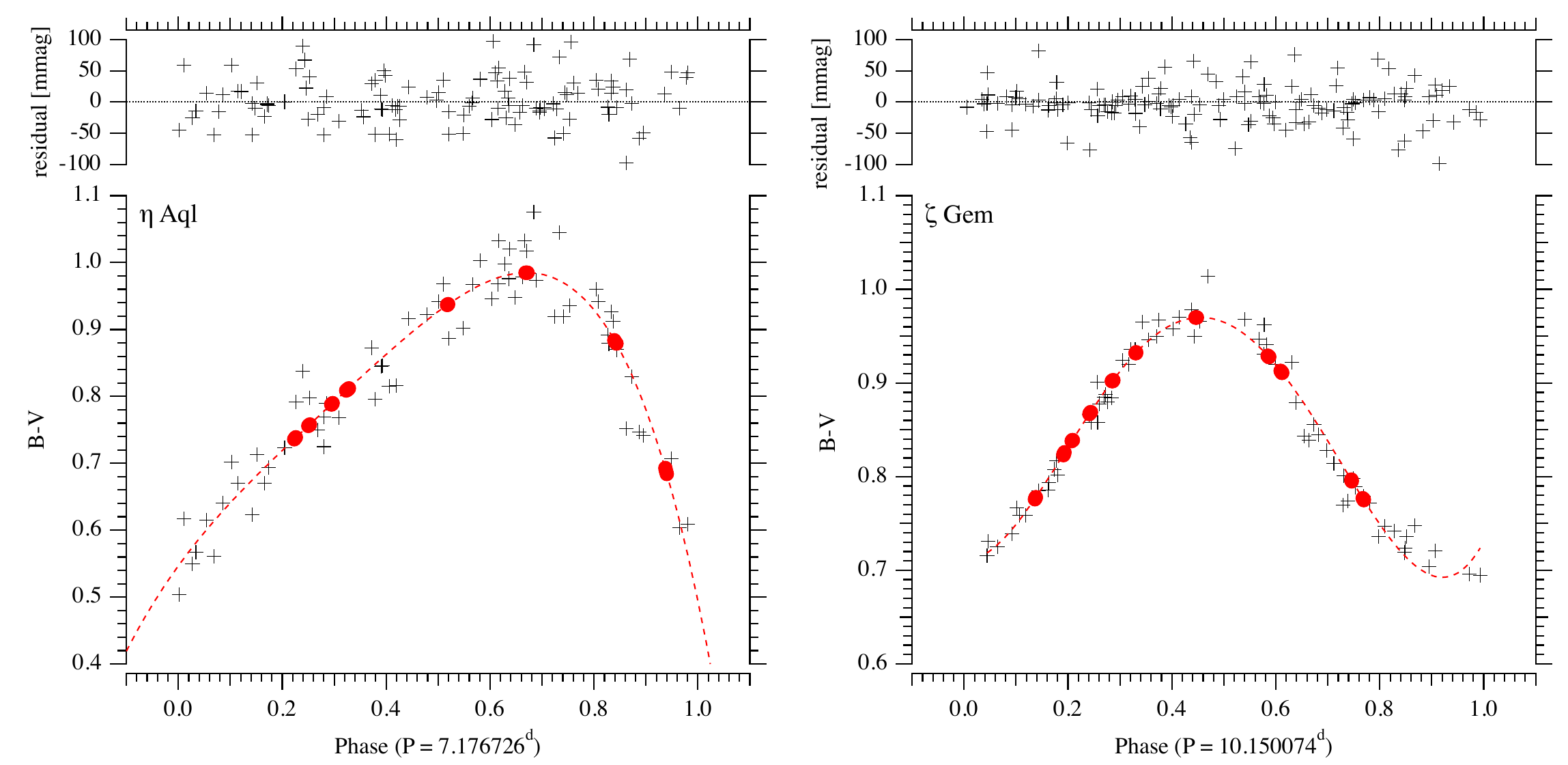}
\caption{$B-V$ variation as a function of phase. Fit is the 5$^{\rm{th}}$ order polynomial listed in Table~\ref{bmvcoefs}. 
The red dots indicate phases at which we secured \FGS astrometry. This fit provides the $B-V$ values used for \eAs and \zGs in Equations 4 and 5.} \label{bmvph}
\end{figure}

The success of single-field parallax astrometry depends on prior knowledge of the reference stars, and sometimes, but less ideally, of the science target. Catalog proper motions with associated errors, lateral color corrections, and estimates for reference star parallax are entered into the modeling as quasi-Bayesian priors, data with which to inform the final solved-for parameters. These values are not entered as hardwired quantities known to infinite precision. We include them as observations with associated errors. The model adjusts the corresponding parameter values within limits defined by the data input errors to minimize $\chi^2$, yielding the most accurate parallax and proper motion for each Cepheid, and in the case of \eA, the best opportunity to measure any reflex motion due to the companion, \eA\,B. We list the various priors below. 
\begin{enumerate} 
\item \textbf{Reference Star Absolute Parallaxes-} Because we measure the parallax of a Cepheid  with respect to 
reference stars which have their own parallaxes, we require estimates of 
 the absolute parallaxes of the reference frame stars.  For past investigations, \cite[e.g.][section 4.1.1]{Ben17}, the colors, spectral type, and luminosity class of a star were used to estimate a spectrophotometric parallax, 
absolute magnitude, $M_V$, and $V$-band absorption, $A_V$. Our task becomes significantly simpler, thanks to \G. The \Gs Early Data Release 3 (EDR3) catalog \citep{Gia21, Lin21a} provides the necessary reference star parallax information with precision and accuracy far superior to that provided by our past determination methodology. Table~\ref{tbl-1} lists the parallax priors used in our modeling, along with \Gs EDR3 ID numbers, \Gs $G$ magnitudes, and the $RUWE$ (Renormalised Unit Weight Error) for \eA, \zGs and each reference star. \cite{Sta21} find that the \Gs $RUWE$  robustly predicts unmodeled photocenter motion, even in the nominal "good" range of 1.0--1.4 \citep[see also][]{Bel20}. Reference star $RUWE$ values (Table~\ref{tbl-1}) suggest  clean reference frames with the \eAs reference star set slightly better. The average reference star parallax error is $\langle \varpi \rangle=0.02$ mas.

\item \textbf{Proper Motions-} We use proper motion priors from the EDR3 with errors on order 
0.05 mas yr$^{-1}$.

\item \textbf{Lateral Color and Cross-Filter Corrections-}  \textrm{These are necessary because the \FGS contains refractive optics, and a neutral density filter required to observe \eA, $V=3.9$,  \zG, $V=3.8$, and \zGs reference star, ref-8 with $V=7.55$. We use  values for those priors from \cite{Ben07}. We list $B-V$ colors used for the lateral color correction in Table~\ref{tbl-IR}, which, for completeness, also contains near-IR colors from 2MASS.}

\item \textbf{Cepheid ($B-V$) vs Phase-}  \textrm{Cepheids exhibit substantial changes in ($B-V$) color index as a function of phase, with $\Delta(B-V) \simeq 0.6$ mag for \eA, and $\Delta(B-V) \simeq 0.3$ for \zG.
To insure the best possible astrometric results in the presence of a lateral color effect, we model sets of ($B-V$)  \cite[][for \eA]{Eng15}  \cite[][for \zG]{Mof80,Ber08} with a fifth-order polynomial (Figure~\ref{bmvph}) 
to predict ($B-V$) at each epoch of \FGS observation. We list the polynomial coefficients in Table~\ref{bmvcoefs}. }

\end{enumerate}

\subsection{The Astrometric Model} \label{AM}
From  
positional measurements we determine for
each observation set  rotation, scale,  and offset parameters relative to an arbitrarily adopted constraint epoch. We employ GaussFit (Jefferys \etal 
1988) 
\nocite{Jef88} to minimize $\chi^2$. The solved equations of condition for are:
\begin{equation}
        x^\prime = x + lc_x(\it B-V)  
\end{equation}
\begin{equation}
        y^\prime = y + lc_y(\it B-V) 
\end{equation}
\textrm{
\begin{equation}
\xi = {A}x^\prime + {B}y' + {C}  - \mu_\alpha \Delta t  - P_\alpha\varpi ~[{- \Delta {\rm XFx}}]
\end{equation}
\begin{equation}
\eta = {D}x^\prime + {E}y^\prime + {F}  - \mu_\delta \Delta t  - P_\delta\varpi ~[{- \Delta{\rm XFy}}]
\end{equation}
}
\noindent 
Identifying terms, $\it x$ and $\it y$ are the measured coordinates from {\it HST};   $(B-V)$ is the Johnson $(B-V)$ color of each star (Table~\ref{tbl-IR});  and $\it lc_x$ and $\it lc_y$ are the lateral color corrections.  \textrm{$ {A}$, $ {B}$, ${D}$, and ${E}$ are scale and rotation plate parameters, ${C}$ and $ {F}$ are offsets}; $\mu_\alpha$ and $\mu_\delta$ are proper motions; $\Delta t$ is the time difference from the constraint epoch; $P_\alpha$ and $P_\delta$ are parallax factors;  and $\it \varpi$ is  the parallax.   
We obtain the parallax factors from a JPL Earth orbit predictor 
(Standish 1990)\nocite{Sta90}, version DE405. $\Delta$XFx and $\Delta$XFy are the cross filter corrections in $\it x$ and $\it y$.
The terms in square brackets, [...] are determined only by the Cepheid, and in the case of \zG, the bright reference star, ref-8.

\subsection{Four Applications of the Model}
Carrying out the following analyses for both Cepheids, we first use their associated reference stars with \Gs EDR3 parallax and proper motion priors (from Table~\ref{tbl-1}) to model scale, rotation, offsets, and individual reference star parallax and proper motion values required to transform the separate observation epochs onto the \Gs reference frame (Step 1). We next apply the derived scale, rotation, and offset values to the Cepheid \FGS observations, deriving only Cepheid parallax and proper motion (Step 2). Step 3 involves re-deriving scale, rotation, and offset values, reference star and Cepheid parallax and proper motion, allowing the Cepheid astrometry to inform the scale, rotation, and offset values. Neither Step 2 nor Step 3 incorporates \Gs EDR3 priors for \eAs or \zG. Step 4 repeats Step 3, this time introducing \Gs EDR3 parallax and proper motion priors for each Cepheid. 
In each Step 2--4 we compare our derived 
Cepheid parallax and proper motion with the EDR3 values listed in Table~\ref{tbl-1}.

 \subsubsection{Step 1: Model the Reference Stars}
\begin{figure}
\includegraphics[width=7in]{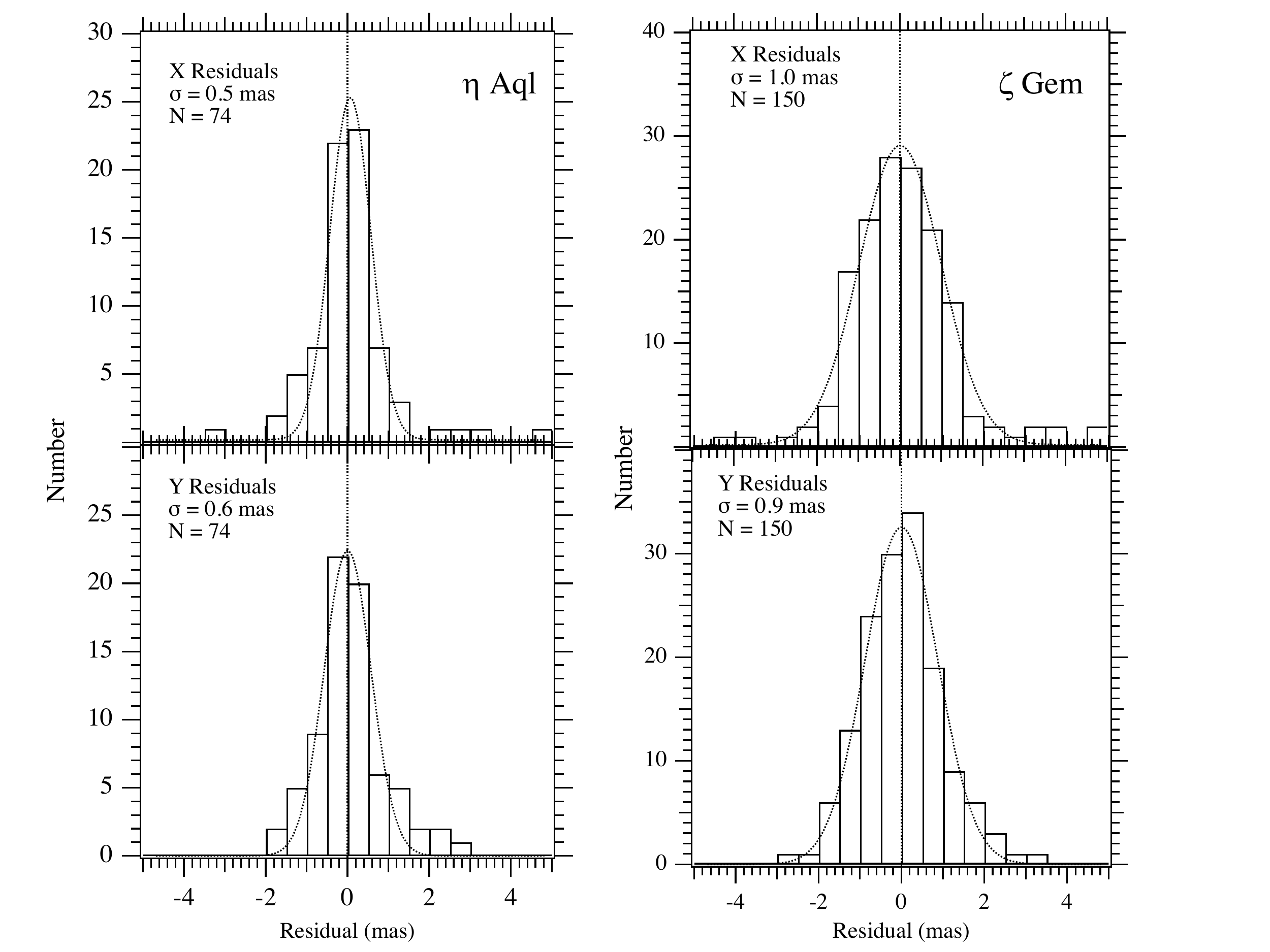}
\caption{Results from Step 1. Histograms of x and y residuals obtained from modeling the \FGS 
observations of  each \FGS reference frame (for \eAs stars 2, 3, 4, 6; for \zGs stars 2, 5, 8, 10, 11; in Figure~\ref{fig-Find}) with Equations 2 -- 5. Distributions are 
fit with gaussians with standard deviations, $\sigma$, indicated in each panel. In addition to fewer reference stars, one entire set of \eAs observations was discarded due to anomalously large residuals.
} \label{fig-FGSH}
\end{figure}

\begin{figure}
\includegraphics[width=7in]{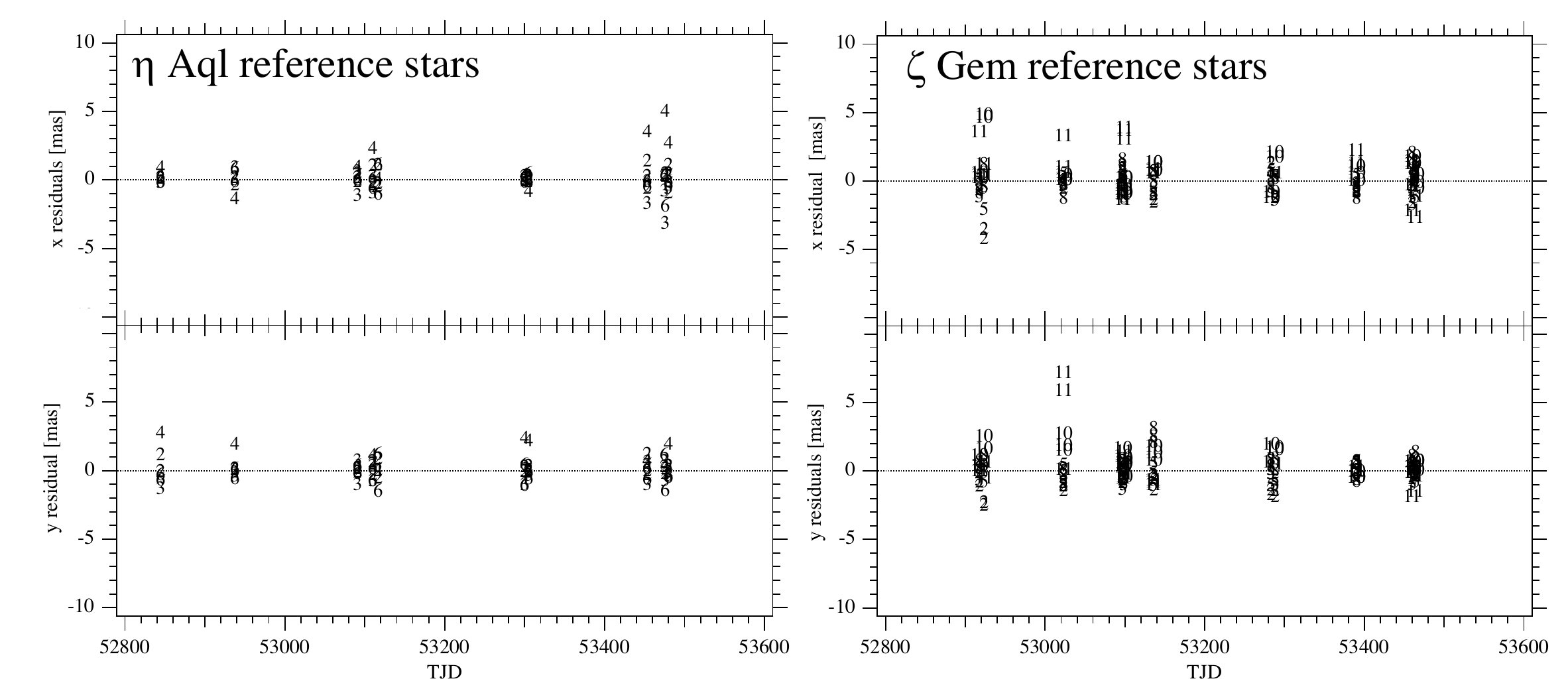}
\caption{Results from Step 1. Reference star position residuals to the Equations 4-7 model  plotted against TJD for the \eAs and \zGs fields.  We identify reference stars by Table~\ref{tbl-1} ID number. Because there are outliers, the equivalent ($x$,$y$) rms values have values,  for \eAs 1.1, 0.8 mas, for \zGs 1.3, 1.2 mas, all larger than the Figure~\ref{fig-FGSH} Gaussian distribution $1-\sigma$ standard deviations for these same residuals.} \label{fig-FGSrest}
\end{figure}

 This step assesses the overall quality of the \FGS astrometry. The Optical Field Angle Distortion (OFAD) calibration \citep{McA02} 
reduces as-built {\it HST} telescope and \FGSns\,1r distortions with magnitude from 
$\sim1\arcsec$ to below 2 mas  over much of the \FGSns\,1r FOV. 
From 
histograms of the \FGS astrometric residuals obtained via Equations 4 to 7 (Figure~\ref{fig-FGSH}) we conclude 
that we have  well-behaved reference star solutions exhibiting residuals with Gaussian distributions with dispersions $\sigma \le 1.0$ mas.   We determine the  \eAs field reference frame 
'catalog' for  \FGSns\,1r in $\xi$ and $\eta$ standard coordinates 
 with average uncertainties, $\langle\sigma_\xi \rangle= 1.2$ and $\langle\sigma_\eta \rangle = 0.5$ mas, again indicating  poorer performance along the \FGSns\,1r $x$ axis. The rms
values for x,y residuals are 1.1, 0.8 mas. Surprisingly, many of the larger residuals come from reference star 4, which has the lowest $RUWE$ value (Table~\ref{tbl-1}). The  \zGs field reference frame 
'catalog' for  \FGSns\,1r in $\xi$ and $\eta$ standard coordinates 
has average uncertainties, $\langle\sigma_\xi \rangle= 0.6$ and $\langle\sigma_\eta \rangle = 0.4$ mas.    For \zGs the rms
values for $x,y$ residuals are 1.3, 1.2 mas. We present reference frame statistics in Table~\ref{tbl-SUM1}. In Figure~\ref{fig-FGSrest} we plot the reference star residuals as a function of time. Neither the \eAs nor the \zGs reference frames exhibit obvious patterns.

\subsubsection{Step 2: Determine Cepheid Parallax and Proper Motion} \label{St2}
Once we have determined plate parameters $A$ through $F$, using only reference star astrometry, they become plate constants in Equations 4 to 7 for modeling, which now includes, in turn, the astrometric measurements for each Cepheid. In other words we rotate, scale, and offset the original Cepheid position measurements into the reference frame defined by their respective reference stars, while solving for Cepheid parallax and proper motion. This model does not include parallax and proper motion priors for the Cepheids. We list these results in Table~\ref{tbl-SUM2}, and display the resulting \eAs and \zGs residuals against TJD in Figure~\ref{fig-Step2}. The \eAs residuals slightly exceed those for \zG. Note for \eAs the significant parallax mismatch compared to the Table~\ref{tbl-1} EDR3 value, and that the \zGs parallax agrees almost perfectly with EDR3.

\subsubsection{Step 3: Cepheid Measurements Contribute to Reference Frame Model while Re-determining Cepheid Parallax and Proper Motion}
For this step the data used to establish the $A-F$ coefficients in Step 1 now include the Cepheid positional measurements, but
without EDR3 parallax and proper motion priors for the Cepheids.  We display the resulting \eAs and \zGs residuals against TJD in Figures~\ref{fig-Step3eA} and \ref{fig-Step3zG}, along with their corresponding  reference star residuals. Note the striking reduction in \eAs residuals compared to the Step 2 result (Figure~\ref{fig-Step2}), this at the expense of inflating (compared to Figure~\ref{fig-FGSrest}) the reference star residuals. This model includes neither parallax nor proper motion priors for the Cepheids, and yields  the parallax and proper motion values listed in Table~\ref{tbl-SUM3}, which for comparison also includes the \cite{Ben07} results for \zG. The \eAs parallax remains significantly different from the EDR3 value, while the \zGs parallax continues to agree within the errors with both the EDR3 and \cite{Ben07} values. 

\subsubsection{Step 4: Re-determine Cepheid Parallax and Proper Motion with Strong EDR3 Priors}
Figures~\ref{fig-Step4eA},~\ref{fig-Step4zG}  and Table~\ref{tbl-SUM4} show the results of including the very restrictive EDR3 priors for parallax and proper motion, again allowing Cepheid astrometry to assist in determining $A-F$. Basically, the \FGS parallaxes of \eAs and \zGs are consistent with the \Gs values, but for \eAs only, by further increasing the \eAs reference star residual rms. These models produce  parallaxes and proper motions that are essentially the input priors, but have decreased the \Gs EDR3 parallax and proper motion formal errors by factors of 2--3. A future, similar re-processing of the other \cite{Ben07} Cepheids might improve the LL in Figure~\ref{fig-LL}.


 \section{Discussion}\label{Disc}
 
 We first review our derived \zGs and \eAs parallaxes, then  discuss possible causes for the large \eAs astrometric residuals obtained from Step 2 (Section~\ref{St2}).
Ascribing the \eAs residual increase to AB system orbital motion, and assuming a B component mass range from the literature, we estimate possible orbit periods and separations. Next, we hypothesize that the \eAs parallax difference between that measured in Step 2 and that predicted  from the \cite{Ben07} LL might result from an A-B system period near one year. Lastly, we investigate the possibility that the observed image motion  is purely a result of the variability of \eA. 

 \subsection{Parallax Results}

We first review the unsurprising \zGs parallax results. No matter what the input priors, the Step 2--4 models yield parallax values (Tables~\ref{tbl-SUM2}, \ref{tbl-SUM3}, \ref{tbl-SUM4}) that all agree within their respective errors, with both the \cite{Ben07} value, and with  EDR3. We interpret this as a validation of our modeling approach.

\begin{figure}
\includegraphics[width=7.0in]{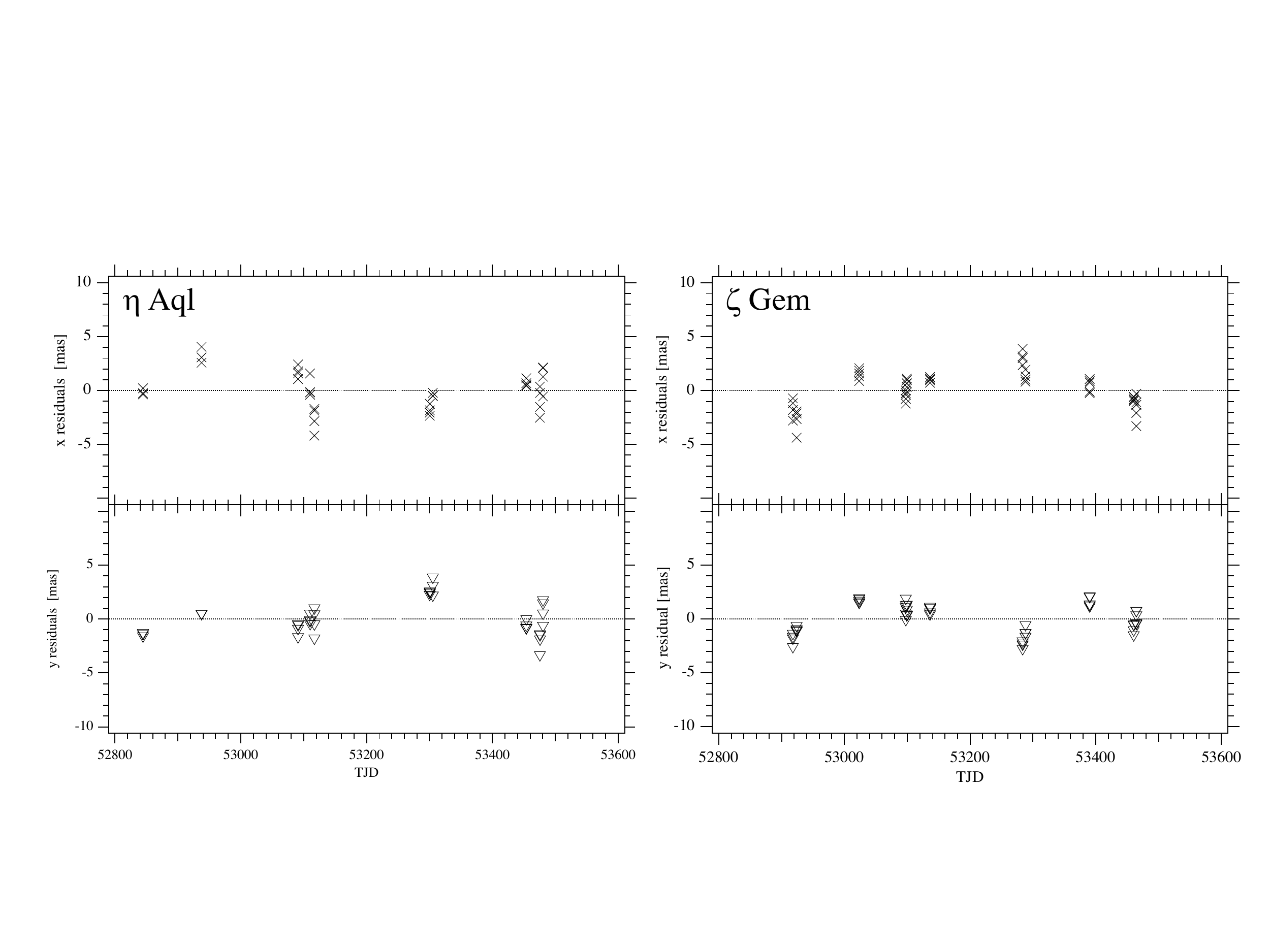}
\caption{Results from Step 2. Cepheid  position residuals plotted against TJD, after applying the Equations 4-7 A-F coefficients while solving for Cepheid parallax and proper motion.   Large residuals suggest unmodeled \eAs motion, with residual rms for \eAs ($x,y$) = (1.8, 1.6 mas). \zG, with no known companion, has rms residuals,  ($x,y$)=(1.7, 1.4 mas). } \label{fig-Step2}
\end{figure}

\begin{figure}
\includegraphics[width=7.0in]{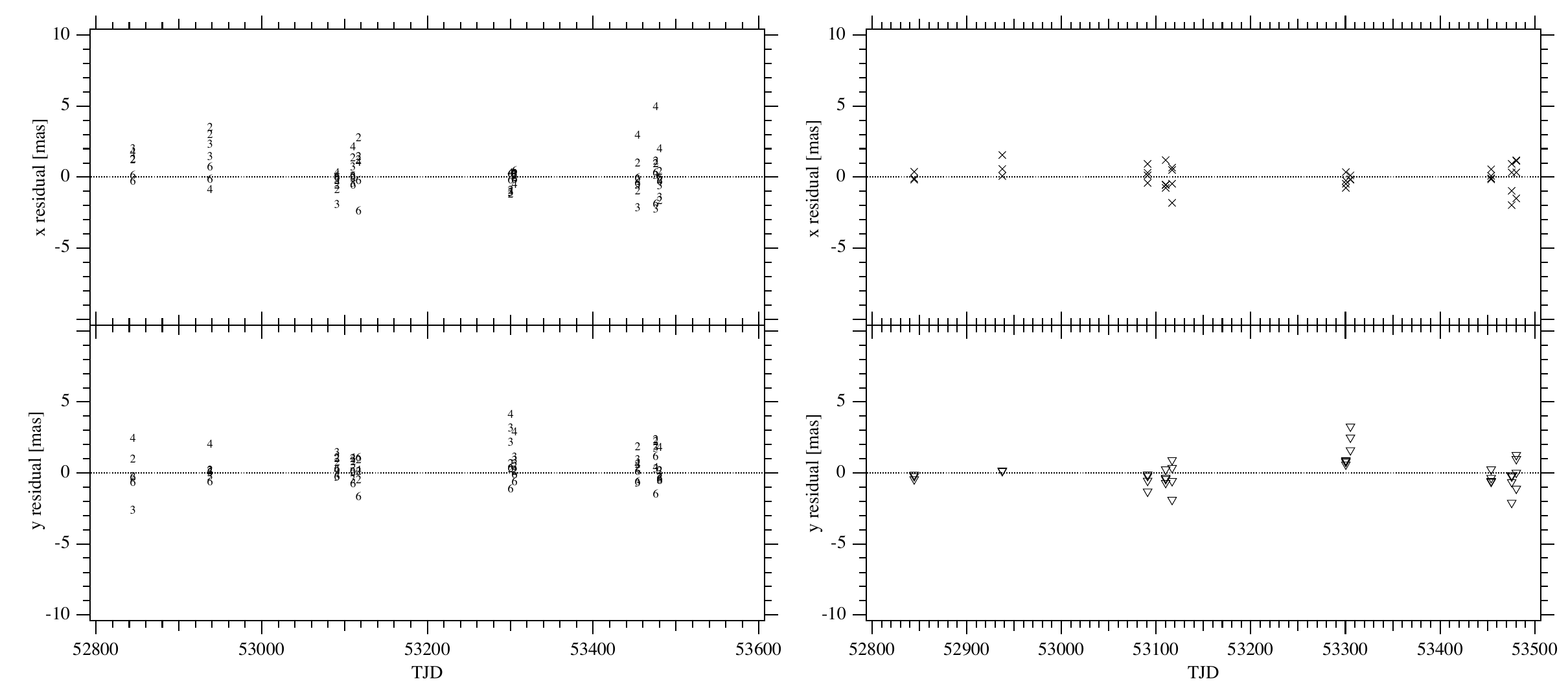}
\caption{ \eAs results from Step 3. Right: \eAs  position residuals plotted against TJD, re-determining the Equations 4-7 A-F coefficients while solving for \eAs parallax and proper motion.   Compared to Figure~\ref{fig-Step2}, allowing \eAs to assist in determining $A-F$ has significantly reduced the \eAs rms residuals from ($x,y$)=(1.8, 1.6 mas) to  (0.8, 1.0 mas). Left: reference star residuals. Compared to Figure~\ref{fig-FGSrest}, including the \eAs measurements has increased the reference star residuals rms from ($x,y$)=(1.1, 0.8 mas) to (1.4, 1.2 mas). These model inputs yield a \eAs parallax, $\varpi=6.55\pm0.25$ mas, which significantly differs from the \Gs EDR3 value, $\varpi=3.67\pm0.19$ mas.} \label{fig-Step3eA}
\end{figure}

\begin{figure}
\includegraphics[width=7.0in]{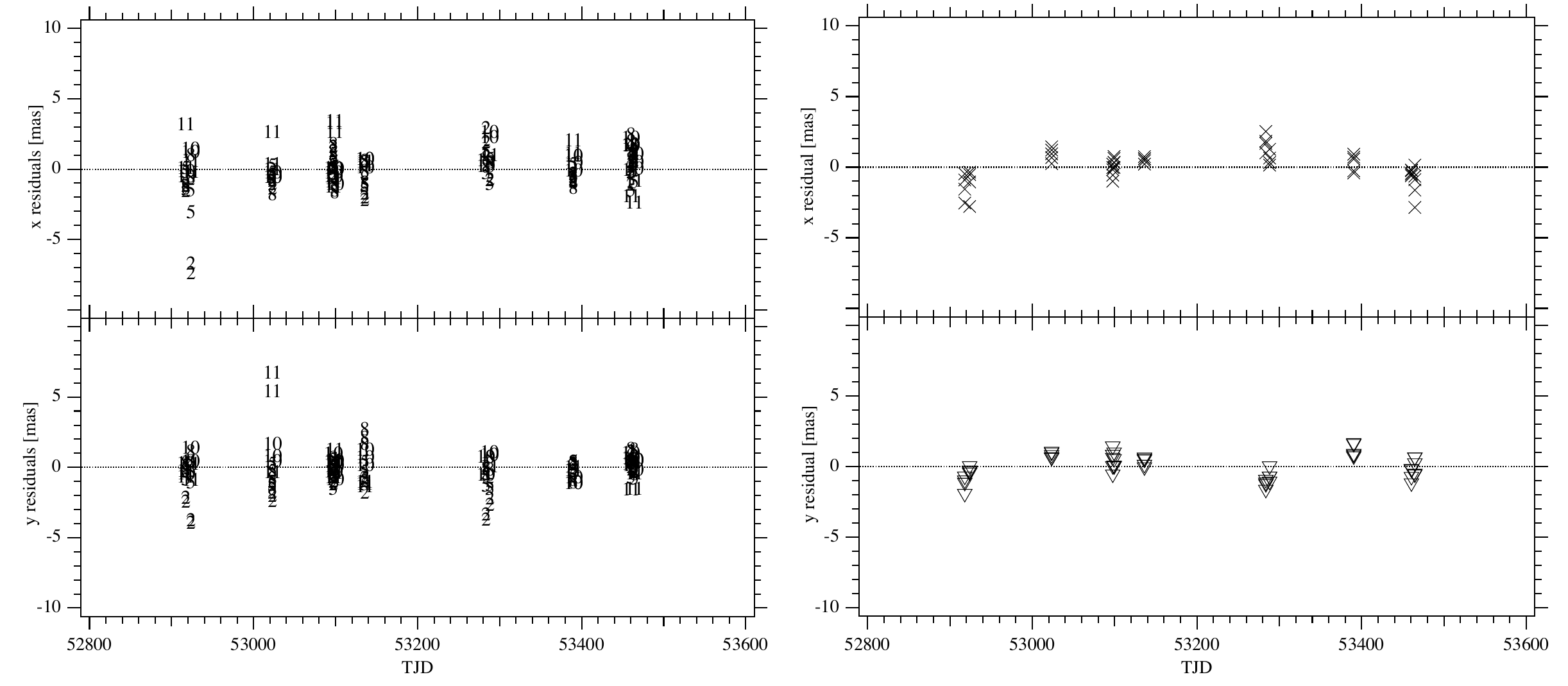}
\caption{\zGs results from Step 3. Right: \zGs  position residuals plotted against TJD, re-determining the Equations 4-7 A-F coefficients while solving for \eAs parallax and proper motion.   Compared to Figure~\ref{fig-Step2}, allowing \zGs to assist in determining $A-F$ has  reduced the \zGs rms residuals from ($x,y$)=(1.8, 1.6 mas) to  (1.5, 1.3 mas), but left the parallax value, $\varpi=3.11\pm0.20$ mas, close to the \Gs EDR3 value. Left: reference star residuals. Compared to Figure~\ref{fig-FGSrest}, allowing the \zGs measurements to inform the $A-F$ coefficients has increased the reference star residuals rms from ($x,y$)=(1.3, 1.2 mas) to (1.5, 1.3 mas). } \label{fig-Step3zG}
\end{figure}

\begin{figure}
\includegraphics[width=7.0in]{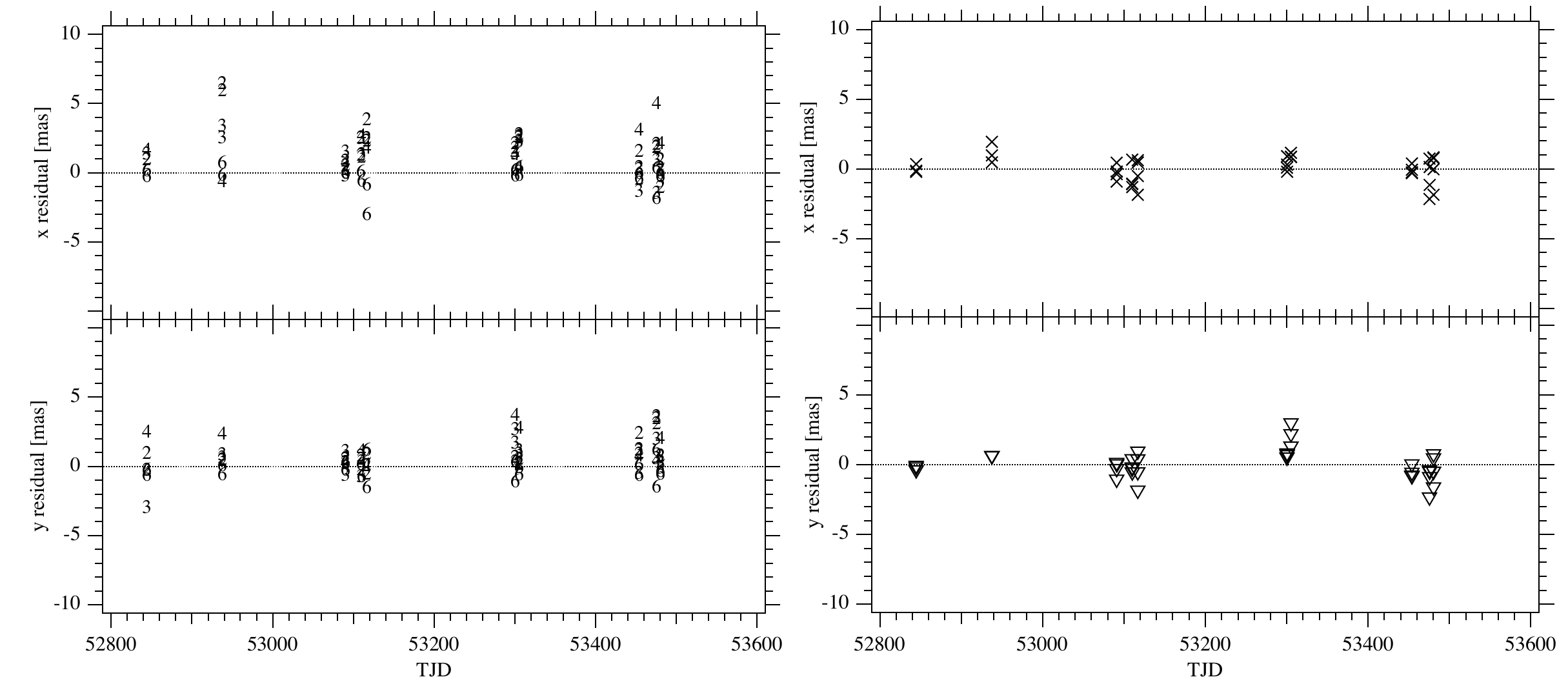}
\caption{\eAs results from Step 4. Right: \eAs  position residuals plotted against TJD, re-determining the Equations 4-7 A-F coefficients while solving for \eAs parallax and proper motion, including EDR3 priors for all stars.   Compared to Figure~\ref{fig-Step2}, allowing \eAs to assist in determining $A-F$ has significantly reduced the rms residuals from ($x,y$)=(1.8, 1.6 mas) to  (0.9, 1.0 mas). Left: reference star residuals. Compared to Figure~\ref{fig-FGSrest}, including the \eAs measurements has significantly increased the reference star residuals rms from ($x,y$)=(1.1, 0.8 mas) to (1.9, 1.3 mas). } \label{fig-Step4eA}
\end{figure}

\begin{figure}
\includegraphics[width=7.0in]{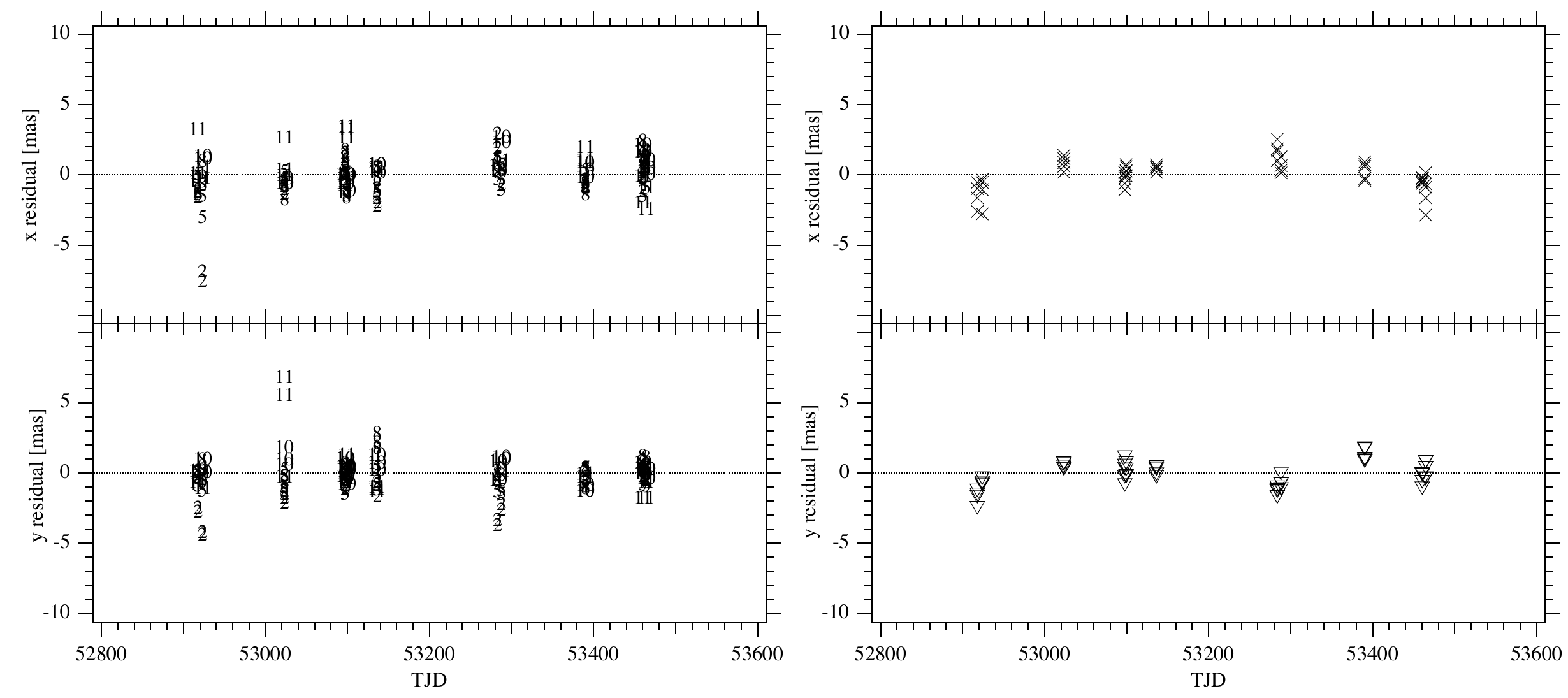}
\caption{\zGs results from Step 4. Right: \zGs  position residuals plotted against TJD, re-determining the Equations 4-7 A-F coefficients while solving for \zGs parallax and proper motion, including EDR3 priors for all stars.   Compared to Figure~\ref{fig-Step2}, allowing \zGs to assist in determining $A-F$ has  reduced the rms residuals from ($x,y$)=(1.8, 1.6 mas) to  (1.1, 0.9 mas). Left: reference star residuals. Compared to Figure~\ref{fig-FGSrest}, including the \zGs priors from EDR3  has increased the reference star residuals rms from ($x,y$)=(1.3, 1.2 mas) to (1.5, 1.3 mas). } \label{fig-Step4zG}
\end{figure}

\begin{figure}
\includegraphics[width=5in]{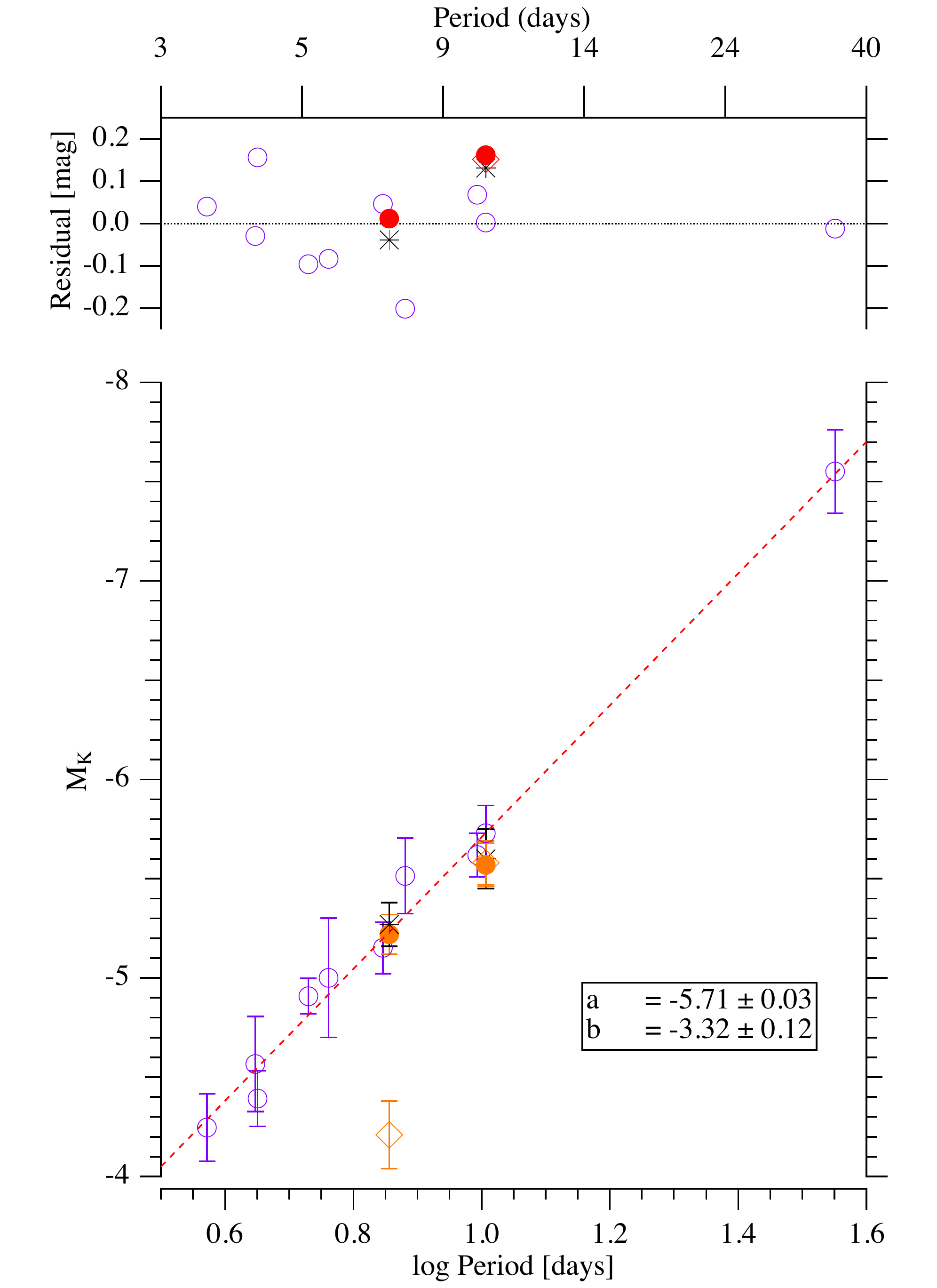}
\caption{The \cite{Ben07} $K-$band LL (\Large \color{blue}$\circ$\color{black}\normalsize) with the Table~\ref{tbl-SUM3} Step 3  (\Large \color{red}$\diamond$\color{black}\normalsize) and Table~\ref{tbl-SUM4} Step 4 (\Large \color{red}$\bullet$\color{black}\normalsize) results for  \eAs  and  \zGs now included at logP values of 0.8559 and 1.0065, respectively. $M_K$ values derived from \Gs EDR3 parallaxes are denoted { \bf(\Large $\ast$\normalsize)}. The linear fit slope and intercept are those previously reported. For \zGs  a model (Step 3) with no EDR3 priors 
produces a parallax and $M_K$ agreeing with the LL. The exact same model with no EDR3 priors applied to \eAs yields a highly discrepant  $M_K$, with a residual ($\Delta M_K=+1.06$) falling outside the residual plot range. 
Including (Step 4) \Gs EDR3 priors for \eAs parallax and proper motion yields agreement with the LL (Table~\ref{tbl-SUM4}). 
\label{fig-LL}}
\end{figure}

From Step 2, as reported in Table~\ref{tbl-SUM2}, applying the Equations 4--7 $A-F$ as plate constants to the measures of \eA , including no EDR3 priors, we obtain a parallax and $K-$band absolute magnitude that place \eAs over one magnitude below the LL determined in \cite{Ben07}.   This model also results in the large \eAs residuals seen in Figure~\ref{fig-Step2}. Step 3, allowing the \eAs measures to contribute to determining the Equations 4--7 $A-F$, while solving for parallax and proper motion, provides no resolution to the parallax disagreement, but does significantly reduce the \eAs residuals, while increasing the reference star residual rms, demonstrating the malleability of the reference frame. Step 4 finally yields a parallax in agreement with EDR3, and that the \eAs reference frame flexes in response to strong Cepheid priors.

\eA\,A, pulsates as a fundamental mode Cepheid with a well-known period. Astrophysical explanations for why it would have an absolute $K-$band magnitude more than    one magnitude lower  than predicted by the \cite{Ben07} LL (Figure~\ref{fig-LL}) do not readily spring to mind. However, could the companion, \eA\,B,  produce a perturbation that could change a measured parallax?

Alternatively, there are a number of differences between the \eAs and \zGs astrometry which might combine to explain the discrepant \eAs parallax: there are only 4 reference stars for \eAs compared to 5 for \zG;
there are 18 observations per reference star for \eAs and 30 observation per reference star for \zG;  there are 37 positional measurements of \eAs compared to  51 for \zG;
there is an additional epoch for \zG; the error along the \FGS x-axis is 6.2 mas for \eA, nearly twice that of \zG. Despite all those differences, both \eAs and \zGs exhibit similar levels of structure in the residuals shown in Figure~\ref{fig-Step2}.

\FGS astrometry of one previous parallax target also yielded  parallax in poor agreement with other determinations. \cite{Ben11} investigated the Population II Cepheid, VY Pyx, finding  a  parallax, $\varpi=6.44\pm0.23$ mas, placing it +1.19 magnitude below a Period-Luminosity Relation defined by five RR Lyr stars and one other Pop II Cepheid, $\kappa$ Pavonis \cite[see figure 3 in][]{Ben17}.  VY Pyx, with a very clean $RUWE=0.89$ value, has a \Gs EDR3 $\varpi=3.95\pm0.02$ mas, placing it on the \cite{Ben11} Period-Luminosity Relation. We   proceed on the assumption that whatever unknown pathology afflicted the VY Pyx data did not similarly impact these \eAs measurements.

\subsection{Searching for \eAs\,B} \label{chk}
Concerning the Step 2 \eAs residuals (Figure~\ref{fig-Step2}, left), could they be evidence of orbital motion? We assert that they are not evidence of modeling issues, given the consistent \zGs parallax results and  the near equality of residual rms values for both \eAs and \zGs at each Step.


Does \Gs provide evidence of anomalous astrometric motion? The \Gs $RUWE$ parameter correlates with photocenter motion \citep{Sta21}. If \eAs\,B and \eA\,A  are a dynamical system, then one might expect an  $RUWE$ value larger than $\sim 1.4$.  \Gs EDR3 catalogs $RUWE$ =2.6 for \eA, consistent with astrometric motion.  However, for \zG,  a Cepheid with no known companion,  $RUWE$ =2.8.  The average $RUWE$ value for the ten Cepheids studied in \cite{Ben07} is $\langle RUWE \rangle = 2.8$. It is unlikely all these Cepheids have astrometrically detectable companions. We ascribe the high  $RUWE$ values to a combination of photometric variability  and that these Cepheids  all have $G<6$, a brightness limit below which requires special and experimental  position extraction.  

We now have access to a second indicator of potential orbital motion,  the \cite{Bra21} $\chi^2$ value. This parameter measures an amount of measured acceleration obtained by comparing an earlier epoch proper motion from \HIP~with a \Gs EDR3 proper motion. A larger $\chi^2$ value indicates more significant  change (acceleration) in proper motion, thus a higher probability of a perturbing companion. The $\chi^2$ values for the Cepheids W Sgr and FF Aql, both confirmed binaries, are 6.03 and 129.2. For \eA, 
$\chi^2$=0.57, lowest of any in the \cite{Ben07} Cepheid list. For \zG, $\chi^2$=1.43. For VY Pyx, our example of previous discrepancy with \G, $\chi^2$=6.70, marginally indicative of unmodeled acceleration.

The question remains; do the relatively large \eAs residuals seen in Figure~\ref{fig-Step2}, indicate unmodeled orbital motion? They do indicate excess motion, when compared to the Figure~\ref{fig-FGSrest} reference star residuals, which are relatively flat-line. 
Comparing \eAs residual rms with reference star residual rms, something perturbs \eAs\,A by 0.7 to 0.8 mas, depending on the axis. As a working hypothesis we assume component B contributes to photocenter motion, thus to
 the large astrometric residuals. 
We identify two possible sources of photocenter motion: orbital and photometric.

\subsection{Photocenter Motion Connected to AB Orbit: Mass and Period Limits for \eAs\,B} \label{Mass}
The RV results yield no period information. They argue for  an \eAs A-B orbit  very close to face-on, or for a very long orbital period. 
In an effort to further constrain information about \eAs\,B we devise and test two hypotheses, assuming a short period: that \eA\,B causes the excess \eA\,A residual noise seen in Step 2 when compared to the Step 1 reference star residuals; that \eA\,B causes the significantly larger than EDR3 \eAs parallax obtained in Step 2 and Step 3.

With these residual rms  excesses, for \eAs\,A  0.7 to 0.8 mas, now hypothesized to be due to orbital motion, we  produce a root sum of squares  perturbation, rss=1.1 mas, which we will use as a constraint on perturbation period, below. Comparing \zGs residual rms with reference star residual rms, we find rms differences of  0.2 to 0.4 mas, depending on the axis. The excess \zGs residual noise seen in Step 2 when compared to the Step 1 reference star residuals yields an rss = 0.44 mas. Presuming no \zGs companion, the  \zGs rss provides only weak evidence for the significance of the \eAs rss signature, which is only 2.5$\times$ as large. 

\begin{figure}
\includegraphics[width=7in]{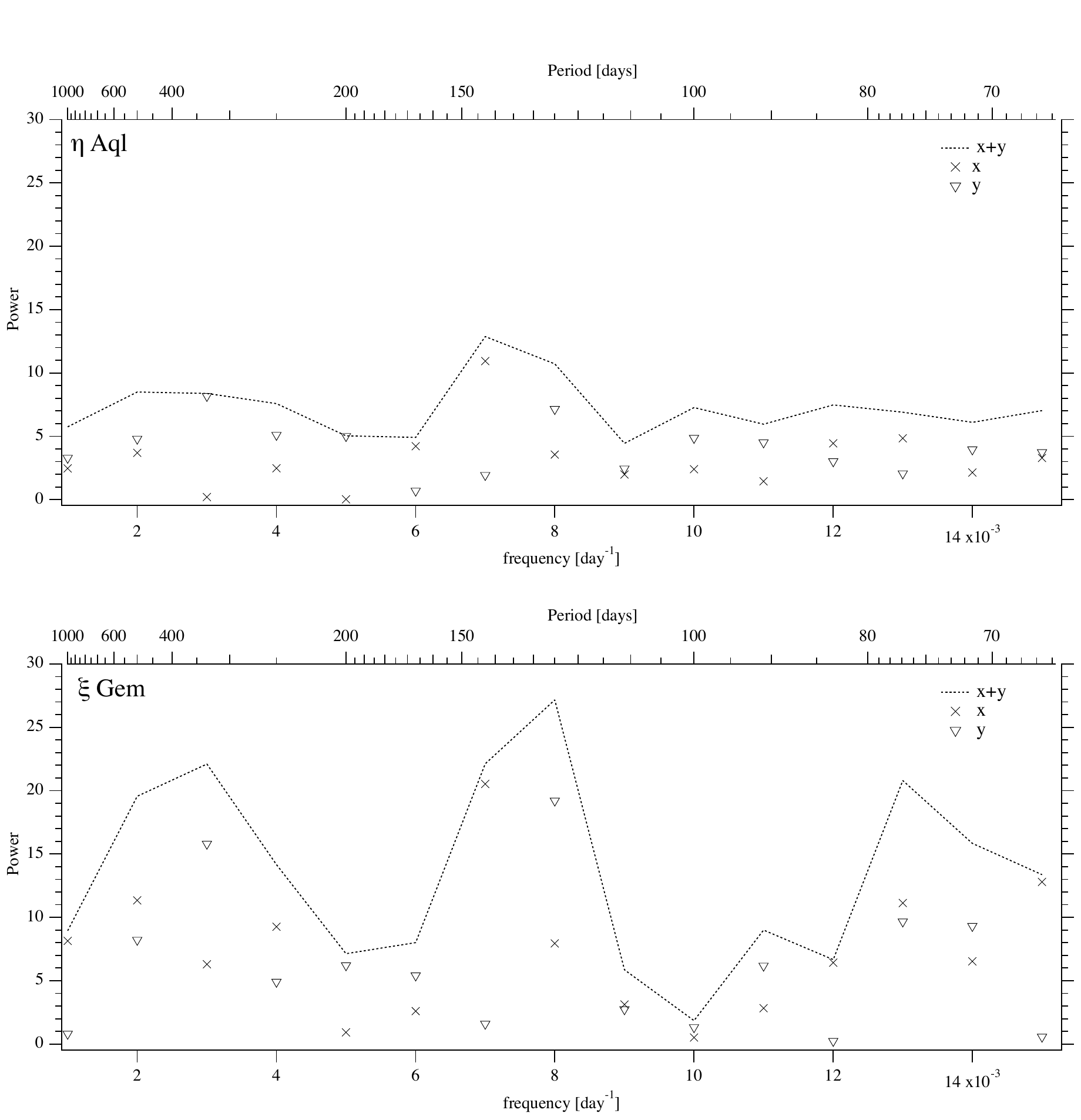}
\caption{ Lomb-Scargle periodogram of the Figure~\ref{fig-Step2} Step 2 residuals. \zGs exhibits significant power (FAP$<<0.001\%$) near one year or  aliases of that period. \eAs residuals exhibit a broad peak near 133$^{\rm d}$ with FAP$\sim0.1\%$.  
\label{fig-LSastr}}
\end{figure}

  Also reducing the effectiveness of this approach, neither Cepheid residual time series evidences unexpected and significant peaks in a periodogram. With effectively five or six observations and relatively uniform spacing (mandated by scheduling at maximum parallax factors, with a few in between),  Lomb-Scargle periodograms (Figure~\ref{fig-LSastr}) of the  astrometric residuals in Figure~\ref{fig-Step2} yield for  \zGs only peaks that are either near one year  or an alias of one year. 
The \eAs residuals exhibit only a broad periodogram peak near one-third year.

\subsubsection{Excess Residual Noise} \label{ERN}
The   Step 2 astrometry provides no \eA\, A-B orbit information other than that of the existence of excess astrometric noise, which we now ascribe to a perturbation caused by \eA\, B.   One might argue that the only valid comparison produced by Step 2
involves differencing the Figure~\ref{fig-Step2} \eAs residual rms (1.8, 1.6 mas) and the \zGs values (1.7, 1.4), for an insignificant  rss = 0.22 mas. The astrometric counter argument: we expect the \zGs residuals to be large because the \zGs reference frame is noisier (Figure~\ref{fig-FGSrest}). Only if the two reference frame noise characteristics were similar (they are not; see Figure~\ref{fig-FGSH}), would a direct \eAs- \zGs comparison yield possible information about \eA\,B.

To establish mass limits for \eA\,B, we appeal to the literature. The established spectral type, B9.8V, is quite close to A0V, a spectral type for which a recent astrometric mass determination exists \citep{Bon17}. They find for Sirius, \m$_{\rm A}=2.06 \pm 0.02 $\msune. Including this we obtain a range of possible B9V-A0V star masses;  $\langle {\cal M} \rangle= 2.5$\msun from binary star astrometry \citep[][]{Tor10a}, $\langle {\cal M} \rangle= 2.9$\msun from stellar models \citep[][]{Aid15}, hence $2.1<\m_{\rm B}<2.9$\msune,  not a particularly tight constraint.

To obtain period limits  for \eAs\,B we utilize the mass function, $f(\m)$,
\begin{equation}
f(\cal{M}) = \frac{\alpha ^{\rm 3}} {\rm P ^2}  = \frac{\cal{M}_{\rm B}^{\rm 3}}  {(\cal{M}_{\rm A} + \cal{M}_{\rm B})^{\rm 2}} \label{Mfrac}
\end{equation}
We next assume a mass for the Cepheid, \m$_{\rm A}$=5.7\msun \citep{Eva13},  our range of possible masses for \eAs\,B, and an estimated perturbation of 1.1 mas from Section~\ref{chk}. We scale the $\alpha$ to AU by adopting the \Gs EDR3 parallax, $\varpi=3.67$ mas.  With these assumptions a 1.1 mas perturbation could be caused by this range of possible masses, $2.9>\m_{\rm B}>2.1$ \msune, with this range of possible periods  $0.30>P_{\rm B}>0.15$ years ($110>P_{\rm B}>55$ days), semi-major axis values ranging $0.89>a>0.58$ AU, and A-B separation range on the sky of $3.0>a>2.1$ mas. We summarize these results in Table~\ref{tbl-MF}. Note that all $a$ values comfortably exceed the interferometrically measured maximum radius of \eA , $R_{\rm max}=0.25$ AU \citep{Mer15}. 
Note that the \Gs EDR3 \eAs parallax value yields an absolute $K-$band magnitude that agrees very well with that predicted by the \cite{Ben07} LL (Figure~\ref{fig-LL}).  The \Gs Observation Forecast Tool (https://gaia.esac.esa.int/gost/) suggests that the average spacing of \eAs measurements included in EDR3 was $\sim22^{\rm d}$. With an estimated period, $P_B\sim 80^{\rm d}$, is it possible that the  31 observational epochs used to produce the EDR3 parallax value have averaged out any perturbations caused by \eA\,B? 

Our derived periods are far shorter than those predicted from the common-envelope evolution studies of \cite{Nei15}, who found it exceedingly unlikely that any Cepheid companion could have $P<1$ yr. However, \eAs is a triple system. Post red giant orbit changes could be possible via the Kozai-Lidov effect, which can shrink an orbit semi-major axis \citep{Nao16}.

\subsubsection{Anomalous Parallax and \eA\,B} \label{APar}

The Step 3 modeling yields a parallax, $\varpi=6.13\pm0.17$ mas, still exceeding the \Gs EDR3 value,  but with  \eAs residuals smaller than seen in Step 2. If the reference frame includes \eA, the \eAs residuals decrease. This casts doubt on the wisdom of appealing to residual rms for perturbation size (Section~\ref{ERN}). Also, our period range violates the \cite{Nei15} limits.
The \eAs Step 3 parallax, $\varpi=6.13\pm0.17$ mas, differs from the EDR3 value, $\varpi=3.67\pm0.19$ mas, by $\Delta \varpi=2.46\pm0.25$ mas. Our second hypothesis, breathtakingly  $ad~ hoc$, supposes that the \eAs A-B orbital period is close to one year, a period, though unlikely, permitted by the \cite{Nei15} results. Hence, that perturbation has an amplitude, $\alpha = 2.46$ mas. 
Our previous loose \eA\,B mass constraint was $2.9>\m_{\rm B}>2.1$ \msune. Equation 8, with $P=1$ year,  \m$_{\rm A}$=5.7\msune, $\alpha = 2.46$ mas, and a scaling parallax, $\varpi=3.67\pm0.19$ mas, yields 
\m$_{\rm B}=1.9\pm0.2$\msune. 
That the EDR3 parallax should have been similarly affected by an AB system one year period argues for the shorter periods discussed in Section~\ref{ERN}.

\subsection{Photometry-induced Image Motion}

\subsubsection{\FGS Response to Non-point Sources} 
 \FGS POS mode works best with point sources. Non-point sources will reduce the amplitude of the interferometric response
 curve \citep[][section 3.5]{Nel12}, decreasing the slope of the response curve, thereby  degrading the positional precision. \eAs varies in size from 1.65 to 1.85 mas as a function of Cepheid pulsational phase with a maximum near Phase=0.4 \citep{Mer15}. \zGs varies from 1.6 to 1.75 mas with a maximum near Phase=0.25 \citep{Bre16}. Inspecting the Step 2 residuals, neither shows an increase in positional scatter at phases of maximum diameter. We conclude that (Cepheid) size doesn't matter. 
\subsubsection{A $\beta$ Constraint?} \label{mobettah}
 \cite{Hei78} defines a luminosity ratio, $\beta$
\beq
\beta= L_{\rm B}/(L_{\rm A} +L_{\rm B}) = 1/(1+10^{0.4\Delta {\rm m}})
\eeq
where L is measured luminosity and $\Delta$m the magnitude difference between components \eA\,A  and \eA\,B,
and a mass fraction, $f$, calculated from the  masses of components A and B
\beq
f=\cal{M}_{\rm B}/(\cal{M}_{\rm B}+\cal{M}_{\rm A})
\eeq
For this purpose we adopt
a distance modulus, $m-M=7.20$ (Table~\ref{tbl-SUM4}), thus an \eAs absolute magnitude range $-2.9<M_V<-3.8$, the variation due to Cepheid pulsation. With \eA\,B $M_V=1.17$ \citep{Eva91}, Cepheid variability produces a variable $\Delta m$, hence a variable  $\beta$, $0.023>\beta>0.010$. (The F1-5 V star, \eA\,C with $M_V\simeq3$, contributes little to  $\beta$.) For the mass fraction
we adopt $f=0.3$ from \mA=5.7\msun and \mB=2.5\msune. At any time Component B is $1-f$ distant from the system center of gravity, and $(1-f)+(f-\beta)$ distant from the center of light (the photocenter). The brighter component A is $\beta$ distant from the photocenter.  Being very small, the changing $\beta$ has very little leverage to change the small separations between A and B hypothesized in Sections~\ref{ERN} and \ref{APar}, hence the measured position of the brighter component, \eA\,A. 

\begin{figure}
\includegraphics[width=7in]{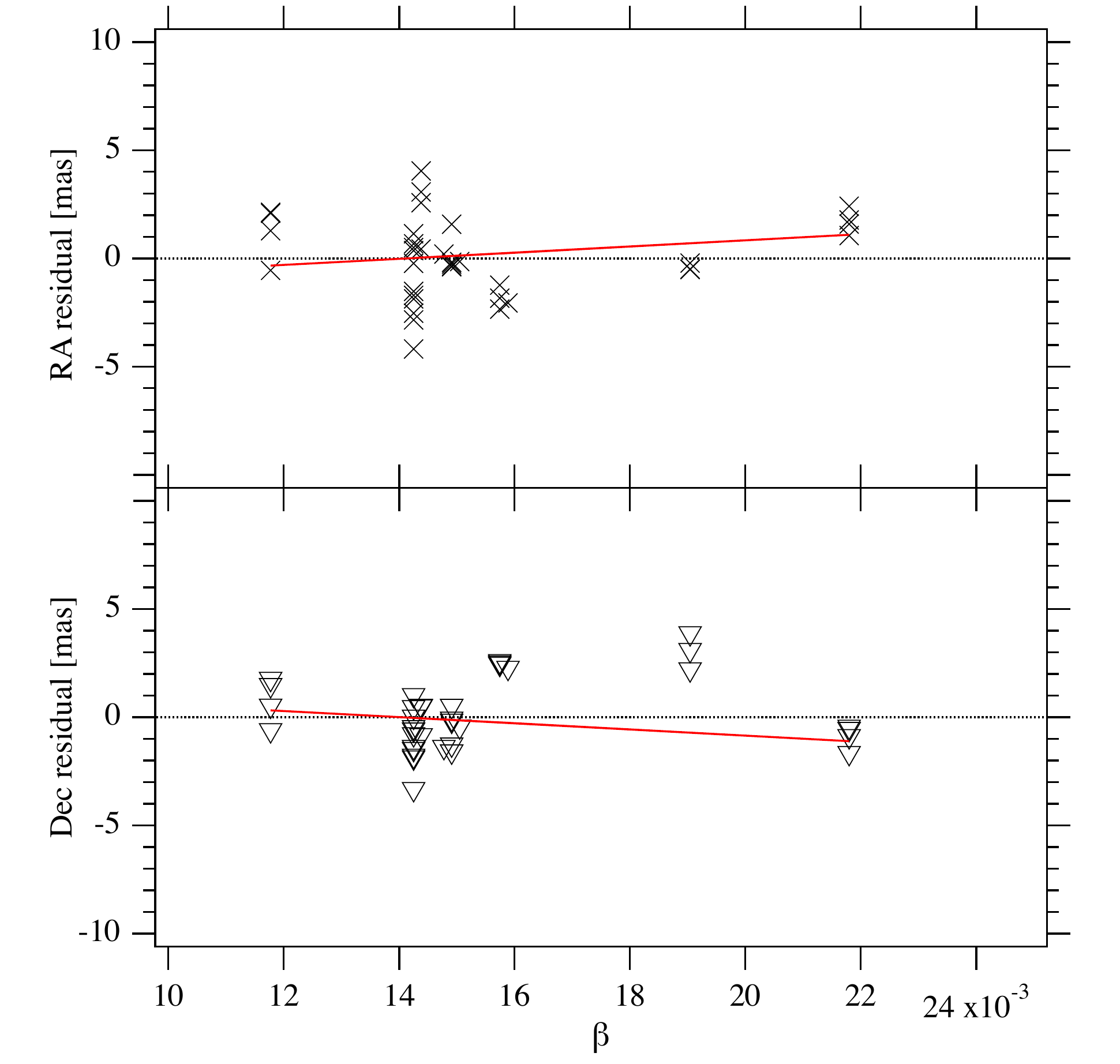}
\caption{ The Step 2 \eA\,A position residuals versus the Section~\ref{mobettah} $\beta$ parameter, showing no strong correlation The lines   linking the smallest and largest measured $\beta$ values (while passing near the average $\beta$=0.015,)  indicate what variation might be caused by  an  \eA\,AB system with separation $\rho_{\rm AB}$= 200 mas at position angle P.A.=45$\arcdeg$.
\label{fig-nobettah}}
\end{figure}

Thus far we treated a photocenter position variation as a nuisance, a source of astrometric noise described through the $\beta$ parameter. We now turn this around and use $\beta$ as a probe. The \eA\,AB system has a constant photometric source, \eA\,B some unknown A-B separation, $\rho_{\rm AB}$, from a known variable source, \eA\,A. The variable  $0.023>\beta>0.010$ has a vanishingly small
effect on  smaller separations (Sections~\ref{ERN} and~\ref{APar}). For large separations, the \eA\,A residuals should correlate with $\beta$, a larger  $\beta$ associated with a larger shift from the average photocenter. Demonstrably, $\beta$ does not strongly correlate with astrometric residual  as shown in Figure~\ref{fig-nobettah}. This lack of correlation supports  small separations.

We now attribute the excess residual (comparing \eAs residuals with the reference star residuals) found in Step 2 only to photocenter variations. We have identified a residual difference 0.7 mas in RA and 0.8 mas in Dec between the \eA\,A position measurements and those of the reference stars. We now hypothesize that this difference is due to $\beta$ alone, working on an unknown  $\rho_{\rm AB}$. A $\rho_{\rm AB}$= 200 mas would cause photocenter motion due only to Cepheid pulsation,  varying between 2.4 and 4.4 mas, a $\Delta\rho_{\rm AB} = \pm1.0$ mas, centered on the average \eAs $\beta=0.015$. This variation is of the same order of magnitude as the excess residuals from Section~\ref{chk}.  A separation, $\rho_{\rm AB}$= 200 mas, and a position angle, P.A.$=45\arcdeg$ produces the lines in Figure~\ref{fig-nobettah}, intersecting the brightest ($\beta=0.012$) and faintest ($\beta=0.022$) phases. The residual pattern in Figure~\ref{fig-nobettah} is consistent with a  separation and position angle, $\rho_{\rm AB}$= 200 mas, P.A.=45$\arcdeg$.  Assuming \m$_{\rm A}$=5.7\msun and \m$_{\rm B}$=2.2\msune, yields $P_{\rm AB}=152$ yr, and undetectable RV variations ($\pm$1 \kms) for inclinations less than $\sim 20\arcdeg$.   These residual=f($\beta$) distributions lack the signal to noise to serve as compelling evidence, but do serve to illustrate a possible technique.
\section{Summary} \label{Summ}

\begin{enumerate}
\item More precise data from the \hets HRS, the Hermes spectrograph on the Mercator Telescope, the Coralie spectrograph on the Euler Telescope, the HJS Telescope, and previously obtained lower-precision measures, yield a Cepheid pulsational RV curve, adequately described by a 12 coefficient Fourier series. 

\item The  lack of any detectable period in the RV residuals obtained by removing the Cepheid RV signature suggests either a nearly face-on orbit or a very long period {\rm for the \eA\,A-B system}.

\item Astrometry (Step 1) of the reference stars associated with  \eAs and \zGs demonstrates 0.6 mas and 1 mas per observation precision respectively.

\item An astrometric re-analysis (Steps 2 through 4) of the \zGs field  yields a parallax agreeing with both \Gs EDR3 and \cite{Ben07}, establishing the robustness of our astrometric modeling.
 
 \item \eAs \hst/\FGS astrometry (Steps 2 and 3) carried out with no prior knowledge of parallax or proper motion resulted in a parallax yielding an absolute $K-$band magnitude  approximately one magnitude  fainter than that predicted by the \cite{Ben07} LL, and with positional residual rms larger than that obtained for the reference stars.

\item Including (Step4) parallax and proper motion priors from \Gs EDR3 {\rm for reference stars and \eAs} resulted in better agreement with EDR3 (with  errors smaller than EDR3) and
{\rm smaller} positional residuals for \eA,   but at the expense of
significantly larger residuals in the reference stars.

\item Neither the \Gs EDR3 $RUWE$ nor the \cite{Bra21} $\chi^2$ values are consistent with astrometric companions for either \eAs or \zG. 

\item Comparing the Step 2 \eAs residuals to the reference star residuals, we determine that \eAs exhibits an rms excess of $\sim1.1$ mas. Assuming the excess comes from orbital motion, assuming for \eA\,A a mass, \m$_{\rm A}$=5.7\msune, and assuming a range of possible masses for
 \eA\,B, $2.9 >$\m$_{\rm B} > 2.1$ \msune, provide a possible period range to $0.15<P_{\rm B}<0.30$ year ($110>P_{\rm B}>55$ days). The \Gs average measurement spacing of $\sim22^{\rm d}$ may have averaged out this perturbation. However, to find a Cepheid companion with this short of a period is highly unlikely, and might require the effects of Kozai-Lidov on the orbit.
 
 \item Hypothesizing that  the parallax mismatch between the Step 3 \FGS result and the \Gs EDR3 result represents a perturbation amplitude from a $P=1$ yr \eA\,A-B orbit suggests \m$_{\rm B}=1.9\pm0.2$\msune. That the EDR3 parallax result is not similarly affected argues against this hypothesis.
 
  \item Ascribing photocenter motion only to the variation of the Cepheid, \eA\,A, in the presence of  the constant brightness companion, \eA\,B, and working only with the brightest and faintest Cepheid phases yields a possible separation, $\rho_{\rm AB}\sim 200$ mas, at a position angle, P.A.$\sim45\arcdeg$, a separation consistent with a long period ($\sim150$ yr) and the observed extremely small RV variation.  The residual=f($\beta$) relations are too noisy to constitute a firm P.A., $\rho$ measurement of an actual binary system.
 
 \item None of these efforts to further characterize  the companion \eA\,B obtained through hypothesis provide any actual \eA\,B orbit information, only results based on conjectures engendered by peculiarities in the astrometric results, which could be previously unidentified systematic errors.

\end{enumerate}

\begin{acknowledgments}

We thank an anonymous (and patient!)  referee for  many useful suggestions that materially improved the organization of this paper. Support for this work (based on observations made with the NASA/ESA Hubble Space Telescope) was provided by NASA through grants GO-09879 and GO-10106 from the Space Telescope Science Institute, which is operated by the Association of Universities for Research in Astronomy (AURA), Inc., under NASA contract NAS5-26555.  NRE acknowledges the Chandra X-ray Center NASA Contract NAS8-03060. RIA acknowledges funding provided by SNSF Eccellenza Professorial Fellowship PCEFP2\_194638. The Hobby-Eberly Telescope spectra could not have been acquired without the dedicated work of the \hets Resident Astronomers ( John Caldwell, Steve Odewahn, Sergey Rostopchin,  Matthew Shetrone) and Telescope Operators (Frank Deglman,  Vicki Riley,  Eusebio Terrazas,  Amy Westfall).  We gratefully acknowledge their contributions. We gratefully acknowledge private communication of $\eta$ Aql velocities from  Joel Eaton, prior to publication, based on spectra from the Tennessee State University Automatic Spectroscopic Telescope.  
This work has made use of data from the European Space Agency (ESA)
mission {\it Gaia} (\url{http://www.cosmos.esa.int/gaia}), processed by
the {\it Gaia} Data Processing and Analysis Consortium (DPAC,
\url{http://www.cosmos.esa.int/web/gaia/dpac/consortium}). Funding
for the DPAC has been provided by national institutions, in particular
the institutions participating in the {\it Gaia} Multilateral Agreement. 
This publication makes use of data products from the Two Micron All Sky Survey, which is a joint project of the University of Massachusetts and the Infrared Processing and Analysis Center/California Institute of Technology, funded by NASA and the NSF. This research has made use of the SIMBAD database, operated at CDS, Strasbourg, France; the NASA / IPAC Extragalactic Database, which is operated by JPL, California Institute of Technology, under contract with NASA; and NASA's Astrophysics Data System Abstract Service.We benefitted from early data reductions by Dr. Jacob Bean.  G.F.B. fondly remembers Debbie Winegarten (R.I.P), whose able assistance with other matters freed me to devote necessary time to this investigation, and thanks the American Astronomical Society, whose support while G.F.B. was AAS Secretary was much appreciated.
\end{acknowledgments}

\bibliography{/Active/myMaster}

\clearpage

\begin{deluxetable}{lllrl}
\tablewidth{0in}
\tablecaption{ \hets Radial Velocities\tablenotemark{a}\label{tbl-HET}}
\tablehead{
\colhead{Template}&
\colhead{TJD\tablenotemark{b}}
&\colhead{Phase}&
\colhead{RV\tablenotemark{c}}&
\colhead{err}
}
\startdata
1&55421.6832&0.392987&14.291&0.037\\
1&55432.6456&0.920442&7.699&0.042\\
1&55432.7468&0.934547&4.856&0.051\\
1&55450.6017&0.422376&14.790&0.039\\
1&55451.5960&0.560923&20.209&0.043\\
1&55458.6719&0.546853&19.075&0.042\\
...&...&...&...&...
\enddata
\tablenotetext{a}{Full table available on-line.}
\tablenotetext{b}{Julian Day - 2400000}
\tablenotetext{c}{Relative velocity in \kms. See Table~\ref{tbl-RVs} for correction to absolute velocity.}
\end{deluxetable}
\begin{deluxetable}{lllrl}
\tablewidth{0in}
\tablecaption{HJS Radial Velocities\tablenotemark{a}\label{tbl-HJS}}
\tablehead{
\colhead{Template}&
\colhead{TJD\tablenotemark{b}}
&\colhead{Phase}&
\colhead{RV\tablenotemark{c}}&
\colhead{err}
}
\startdata
1&58376.5856&0.116495&-7.841&0.184\\
1&58376.5873&0.116735&-7.945&0.171\\
1&58376.5891&0.116979&-7.735&0.199\\
1&58390.6463&0.075647&-9.126&0.170\\
1&58390.6476&0.075830&-9.045&0.172\\
1&58390.6489&0.076014&-8.996&0.178\\
...&...&...&...&...
\enddata
\tablenotetext{a}{Full table available on-line.}
\tablenotetext{b}{Julian Day - 2400000.0}
\tablenotetext{c}{Relative velocity in \kms. See Table~\ref{tbl-RVs} for correction to absolute velocity.}
\end{deluxetable}
\begin{deluxetable}{llrl}
\tablewidth{0in}
\tablecaption{ Eaton Radial Velocities\tablenotemark{a}\label{tbl-EAT}}
\tablehead{
\colhead{TJD\tablenotemark{b}}
&\colhead{Phase}&
\colhead{RV\tablenotemark{c}}&
\colhead{err}
}
\startdata
54386.6087&0.169800&-26.87&0.10\\
54388.6734&0.457487&-15.78&-0.02\\
54391.6386&0.870647&-8.52&0.16\\
54396.6376&0.567188&-10.13&0.33\\
54399.6884&0.992275&-31.76&0.27\\
...&...&...&...
\enddata
\tablenotetext{a}{Full table available on-line.}
\tablenotetext{b}{Julian Day - 2400000.0}
\tablenotetext{c}{Relative velocity in \kms. See Table~\ref{tbl-RVs} for correction to absolute velocity.}
\end{deluxetable}

\begin{deluxetable}{l l r l l}
\tablewidth{5in}
\tablecaption{ Coralie and Hermes Radial Velocities\tablenotemark{a}\label{tbl-RIA}}
\tablehead{
\colhead{TJD\tablenotemark{b}}
&\colhead{Phase}&
\colhead{RV\tablenotemark{c}}&
\colhead{err}&
\colhead{Source\tablenotemark{d}}
}
\startdata
56096.6778&0.443989&-15.55&0.02&H\\
56097.6907&0.585123&-7.74&0.02&H\\
56098.6831&0.723398&6.03&0.02&H\\
56099.6977&0.864769&-6.77&0.02&H\\
56855.5415&0.180871&-25.93&0.02&H\\
56856.5404&0.320044&-19.51&0.02&H\\
...&...&...&...&...
\enddata
\tablenotetext{a}{Full table available on-line.}
\tablenotetext{b}{Julian Day - 2400000.0}
\tablenotetext{c}{Relative velocity in \kms. See Table~\ref{tbl-RVs} for correction to absolute velocity.}
\tablenotetext{d}{H=Hermes spectrograph on the Mercator Telescope; C=Coralie spectrograph on the 
Euler Telescope }
\end{deluxetable}


\begin{deluxetable}{l l l r}
\tablewidth{5in}
\tablecaption{ Sources of \eAs Radial Velocities\label{tbl-RVs}}
\tablehead{
\colhead{Key\tablenotemark{a}} &
\colhead{\# RV} &
\colhead{$\gamma$\tablenotemark{b}}&
\colhead{source}
}
\startdata
Kiss \& Vinko&14&-0.68$\pm$0.11&\cite{Kis00}\\
Bersier&38&-0.36 0.10&\cite{Ber02}\\
Storm&26&-0.23 0.20&\cite{Stor04}\\
TGB05&30&-0.56 0.18&\cite{Bar05}\\
Eaton&217&0.00 ~-&\cite{Eat20}\\
HET T1&34&-30.62 0.03&HET, Template 1, this paper\\
HET T2&36&-30.72 0.04&HET, Template 2, this paper\\
HET T3&25&-21.50 0.05&HET, Template 3, this paper\\
HJS T1&25&-21.53 0.08&HJS, Template 1, this paper\\
HJS T2&26&-4.77 0.09&HJS, Template 2, this paper\\
Borgniet&13&-0.12 0.14&\cite{Bor19}\\
Cor \& Herm&55&-0.56 0.03&Coralie \& Hermes, this paper
\enddata
\tablenotetext{a}{Legend symbol in Figures~\ref{fig-RVph} and \ref{fig-RVt}}
\tablenotetext{b}{Velocity offsets required to minimize residuals to a Fourier description of 
the pulsation velocity of \eA, assuming an Eaton offset, $\gamma=0$.}
\end{deluxetable}
\begin{deluxetable}{l r l r l}
\tablewidth{0in}
\tablecaption{Fourier Coefficients Describing the \eAs Pulsational RV Variation \label{tbl-Fou}}
\tablehead{
\colhead{j}&
\colhead{a}&
\colhead{$\sigma_{\rm a}$}&
\colhead{b} &
\colhead{$\sigma_{\rm b}$}
}
\startdata
0&-15.328&0.021&0.000&0.000\\
1&-7.389&0.019&-13.510&0.016\\
2&-7.876&0.019&-1.626&0.016\\
3&-1.916&0.018&2.445&0.017\\
4&-0.344&0.017&1.394&0.018\\
5&0.794&0.015&0.467&0.020\\
6&0.250&0.015&0.024&0.019\\
7&0.088&0.014&-0.160&0.020\\
8&-0.085&0.016&-0.003&0.017\\
9&-0.134&0.016&0.026&0.017\\
10&-0.082&0.016&0.015&0.016\\
11&0.025&0.016&0.079&0.016\\
12&0.067&0.015&0.017&0.016
\enddata
\end{deluxetable}

\begin{deluxetable}{r l l l l l l}
\tablewidth{0in}
\tablecaption{ \Gs EDR3 Astrometry Priors\tablenotemark{a}\label{tbl-1}}
\tablehead{
\colhead{ID}
&\colhead{EDR3 ID}&
\colhead{$\varpi$}&
\colhead{$\mu_{\alpha}$}&
\colhead{$\mu_{\delta}$}&
\colhead{$RUWE$}&
\colhead{$G$}
}
\startdata
&&&\eA&&&\\
1&Gaia EDR3 4240272953377646592&3.67$\pm$0.19&8.89$\pm$0.18&-8.32$\pm$0.14&2.561&3.748$\pm$0.014\\
2&Gaia EDR3 4240273159535375104&0.24 0.01&-3.73 0.02&-4.38 0.01&1.049&13.258 0.003\\
3&Gaia EDR3 4240272334901581184&0.84 0.03&-9.28 0.03&-3.90 0.03&1.023&15.242 0.003\\
4&Gaia EDR3 4240272644139211904&0.19 0.03&-0.88 0.03&-3.44 0.02&0.952&15.188 0.003\\
6&Gaia EDR3 4240272781578178048&0.45 0.02&-2.74 0.02&-11.40 0.02&1.02&14.250 0.003\\
&&&\zG&&&\\
1&Gaia EDR3 3366754155291545344&3.07$\pm$0.22&-7.74$\pm$0.25&-0.94$\pm$0.17&2.778&3.540$\pm$0.006\\
2&Gaia EDR3 3366753296297433984&0.37 0.02&-0.86 0.02&-2.71 0.01&0.979&13.589 0.003\\
5&Gaia EDR3 3366795558775643904&1.26 0.04&-4.83 0.05&-7.39 0.03&2.170&12.267 0.003\\
8&Gaia EDR3 3366754464528540416&28.64 0.02&-81.97 0.03&41.33 0.02&0.848&7.432 0.003\\
10&Gaia EDR3 3366754086571429376&0.48 0.02&2.19 0.02&-0.03 0.02&1.036&14.291 0.003\\
11&Gaia EDR3 3366754395807337600&0.57 0.01&4.16 0.01&-4.49 0.01&1.027&12.336 0.003
\enddata
\tablenotetext{a}{Units are: $\varpi$, mas; $\mu_{\alpha}, \mu_{\delta}$, mas yr$^{-1}$. In each field ID=1 denotes the Cepheid.}
\end{deluxetable}

\begin{deluxetable}{ccccccc}
\tablewidth{0in}
\tablecaption{Visible and Near-IR\tablenotemark{a} Photometry \label{tbl-IR}}
\tablehead{\colhead{ID}&
\colhead{$V$} &
\colhead{$B-V$} &
\colhead{$K$} &
\colhead{$(J-H)$} &
\colhead{$(J-K)$} &
\colhead{$(V-K)$} 
}
\startdata
&&&\eA&&&\\
1&3.91 $\pm$ -&0.80$\pm$ 0.03&1.98 $\pm$0.01&0.40$\pm$ 0.01&0.48$\pm$ 0.01&1.93  $\pm$-\\
2&13.68 0.03&1.53 0.03&10.142 0.025&0.796 0.033&0.927 0.036&3.778 0.04\\
3&15.25 0.03&0.75 0.05&13.453 0.031&0.454 0.046&0.449 0.048&1.847 0.04\\
4&15.31 0.03&1.13 0.04&12.919 0.035&0.529 0.040&0.629 0.045&2.451 0.05\\
6&14.40 0.03&1.02 0.05&11.905 0.026&0.629 0.036&0.706 0.034&2.645 0.04\\
&&&\zG&&&\\
1&3.79$\pm$ -& $\pm$ -&2.182$\pm$ 0.288&0.401$\pm$ 0.359&0.257$\pm$ 0.387&1.858$\pm$ -\\
2&13.78 0.06&1.74 0.1&11.537 0.018&0.567 0.033&0.614 0.028&2.303 0.06\\
5&12.36 0.03&0.74 0.06&11.103 0.020&0.307 0.030&0.331 0.028&1.267 0.04\\
8&7.55 0.02&0.69 0.05&6.108 0.023&0.366 0.028&0.373 0.030&1.452 0.03\\
10&14.25 0.02&0.61 0.04&13.127 0.032&0.302 0.044&0.348 0.041&1.203 0.04\\
11\tablenotemark{b}&12.56 0.03&0.66 0.1&  -  -&   - -&   - -&   - -
\enddata
\tablenotetext{a}{Sources: for \eAs and \zGs(ID=1), \cite{Bar97,Wel84}; for reference stars (ID=2-6, =2-11), 2Mass \citep{Skr06} and sources discussed in \cite{Ben07}.}
\tablenotetext{b}{Star too close to \zGs for 2Mass measurement.}
\end{deluxetable}

\begin{deluxetable}{ccc}
\tablewidth{0in}
\tablecaption{$B-V$\tablenotemark{a} as a 5$^{\rm{th}}$ Order Polynomial Function of Phase  \label{bmvcoefs}}
\tablehead{\colhead{term}& \colhead{Value}&
\colhead{error} } 
\startdata
&{\underline \eA}&\\
K0	&0.5468 & 0.0226\\
K1	&1.0729 & 0.332\\
K2	&-1.6424 & 1.41\\
K3	&3.553 & 2.19\\
K4	&-3.0371 & 1.11\\
&{\underline \zG}&\\
K0&0.7102&	0.0150\\
K1	&-0.0111&	0.1871\\
K2&	5.0592	&0.7227\\
K3&	-11.0439	&1.0546\\
K4&	6.0166	&0.5133\\
\enddata 
\tablenotetext{a}{Figure~\ref{bmvph}.}
\end{deluxetable}


\begin{deluxetable}{ll}
\tablecaption{Reference Frame Statistics\label{tbl-SUM1}}
\tablewidth{3in}
\tablehead{\colhead{Parameter} &  \colhead{Value} }
\startdata
~~~~~~~~~~~~~~~~~~ {\underline \eA} &\\
Study duration  &1.74 y  \\
number of observation sets    &   10  \\
reference star average $V$  &  14.66     \\
reference star average $(B-V)$  &1.11   \\
field $A_K$ & 0.05 mag\\
$x$ residual rms&1.1 mas\\
$y$ residual rms&0.8 mas\\
~~~~~~~~~~~~~~~~~~{\underline \zG}&\\
Study duration  &1.50 y  \\
number of observation sets    &   11  \\
reference star average $V$ &  12.03     \\
reference star average $(B-V)$ &0.69   \\
field $A_K$ & 0.00 mag\\
$x$ residual rms&1.3 mas\\
$y$ residual rms&1.2 mas\\
\enddata
\end{deluxetable}


\begin{deluxetable}{ll}
\tablecaption{Step 2: Cepheid Parallax, Proper Motion, and Absolute Magnitude\label{tbl-SUM2}}
\tablewidth{3in}
\tablehead{\colhead{Parameter} &  \colhead{Value} }
\startdata
~~~~~~~~~~~~~~~~~~ {\underline \eA} &\\
$\varpi$& 6.55 $\pm$ 0.25    mas \\ 
$\mu_{\alpha}$ & 7.19 $\pm$ 0.25 mas  yr$^{-1}$\\
$\mu_{\delta}$ & -7.50 $\pm$ 0.27 mas  yr$^{-1}$\\
$\vec{\mu}$ & 10.39 mas  yr$^{-1}$\\
P.A. & $136\fdg2$\\
LKH corr. & -0.01 mag\\
field $A_K$ & 0.05 mag\\
$(m-M)_0$ & 5.97 mag\\
$M_K$ & -4.01 $\pm$ 0.08 mag\\
~~~~~~~~~~~~~~~~~~ {\underline \zG} &\\
$\varpi$& 3.00 $\pm$ 0.32    mas \\ 
$\mu_{\alpha}$ & -7.51 $\pm$ 0.49 mas  yr$^{-1}$\\
$\mu_{\delta}$ & -0.19 $\pm$ 0.29 mas  yr$^{-1}$\\
$\vec{\mu}$ & 7.51 mas  yr$^{-1}$\\
P.A. & $268\fdg6$\\
LKH corr. & -0.09 mag\\
field $A_K$ & 0.02 mag\\
$(m-M)_0$ & 7.70 mag\\
$M_K$ & -5.71 $\pm$ 0.23 mag\\
\enddata
\end{deluxetable}


\begin{deluxetable}{ll}
\tablecaption{Step 3: Cepheid Parallax, Proper Motion, and Absolute Magnitude\label{tbl-SUM3}}
\tablewidth{3in}
\tablehead{\colhead{Parameter} &  \colhead{Value} }
\startdata
~~~~~~~~~~~~~~~~~~ {\underline \eA} &\\
$\varpi$& 6.13 $\pm$ 0.17    mas \\ 
$\mu_{\alpha}$ & 7.62 $\pm$ 0.17 mas  yr$^{-1}$\\
$\mu_{\delta}$ & -7.39 $\pm$ 0.20 mas  yr$^{-1}$\\
$\vec{\mu}$ & 10.62 mas  yr$^{-1}$\\
P.A. & $134\fdg$1\\
LKH corr. & -0.01 mag\\
field $A_K$ & 0.05 mag\\
$(m-M)_0$ & 6.11 mag\\
$M_K$ & -4.15 $\pm$ 0.06mag\\
~~~~~~~~~~~~~~~~~~ {\underline \zG} &\\
$\varpi$& 3.11 $\pm$ 0.20   mas \\ 
$\mu_{\alpha}$ & -7.58 $\pm$ 0.31 mas  yr$^{-1}$\\
$\mu_{\delta}$ & -0.26 $\pm$ 0.18 mas  yr$^{-1}$\\
$\vec{\mu}$ & 7.58 mas  yr$^{-1}$\\
P.A. & $268\fdg1$\\
LKH corr. & -0.03 mag\\
field $A_K$ & 0.02 mag\\
$(m-M)_0$ & 7.57 mag\\
$M_K$ & -5.58$\pm$ 0.14mag\\
~~~~~~~~~~~~~~~~~~ {\underline \zG \tablenotemark{a}} &\\
$\varpi$& 2.78 $\pm$ 0.18    mas \\
$\mu_{\alpha}$ & -6.18 $\pm$ 0.15 mas  yr$^{-1}$\\
$\mu_{\delta}$ & +0.20 $\pm$ 0.21 mas  yr$^{-1}$\\
$\vec{\mu}$ & 6.2  mas  yr$^{-1}$\\
P.A. & 272 $\arcdeg$\\
LKH corr. & -0.03 mag\\
field $A_K$ & 0.02 mag\\
$(m-M)_0$ & 7.81 mag\\
$M_K$ & -5.73 $\pm$ 0.14mag\\
\enddata
\tablenotetext{a}{Result from \cite{Ben07}, using parallax priors derived spectrophotometrically and proper motion priors from UCAC2 \citep{Zac04} }.
\end{deluxetable}


\begin{deluxetable}{ll}
\tablecaption{Step 4: Parallax, Proper Motion, and Absolute Magnitude\label{tbl-SUM4}}
\tablewidth{3in}
\tablehead{\colhead{Parameter} &  \colhead{Value} }
\startdata
~~~~~~~~~~~~~~~~~~ {\underline \eA} &\\
$\varpi$& 3.71 $\pm$ 0.07    mas \\ 
$\mu_{\alpha}$ & 8.89 $\pm$ 0.05 mas  yr$^{-1}$\\
$\mu_{\delta}$ & -8.31 $\pm$ 0.06 mas  yr$^{-1}$\\
$\vec{\mu}$ & 12.16 mas  yr$^{-1}$\\
P.A. & $133\fdg$1\\
LKH corr. & -0.00 mag\\
field $A_K$ & 0.05 mag\\
$(m-M)_0$ & 7.20 mag\\
$M_K$ & -5.22 $\pm$ 0.04mag\\
~~~~~~~~~~~~~~~~~~ {\underline \zG} &\\
$\varpi$& 3.08 $\pm$ 0.06   mas \\ 
$\mu_{\alpha}$ & -7.73 $\pm$ 0.08 mas  yr$^{-1}$\\
$\mu_{\delta}$ & -0.91 $\pm$ 0.06 mas  yr$^{-1}$\\
$\vec{\mu}$ & 7.79 mas  yr$^{-1}$\\
P.A. & $263\fdg3$\\
LKH corr. & -0.00 mag\\
field $A_K$ & 0.02 mag\\
$(m-M)_0$ & 7.56 mag\\
$M_K$ & -5.57$\pm$ 0.04mag\\
\enddata
\end{deluxetable}

\begin{deluxetable}{c c c c c c c c c}
\tablewidth{0in}
\tablecaption{\eAs A-B Period Range from \eA\,B Mass Range \label{tbl-MF}}
\tablehead{
\colhead{\m$_{\rm A}$} &
\colhead{\m$_{\rm B}$} &
\colhead{$P$ [yr]} &
\colhead{$P$ [d]} &
\colhead{$a$ [AU]} &
\colhead{$\alpha$ [AU]} &
\colhead{$\alpha$ [mas]}&
\colhead{$a$ ["]}
}
\startdata
5.7&2.1&0.27&97&0.82&0.3&1.1&0.0030\\
5.7&2.2&0.25&91&0.79&0.3&1.1&0.0029\\
5.7&2.3&0.23&84&0.75&0.3&1.1&0.0028\\
5.7&2.7&0.18&66&0.65&0.3&1.1&0.0024\\
5.7&2.9&0.15&55&0.58&0.3&1.1&0.0021
\enddata
\end{deluxetable}

%
%
\clearpage

\clearpage

\end{document}